\documentclass[11pt, letterpaper]{article} 

\usepackage[includeheadfoot,
            marginratio={1:1,2:3}, 
            width=440pt, 
            height=688pt,]{geometry}

\usepackage{amsmath}
\usepackage{amsfonts}
\usepackage{amssymb}
\usepackage{bbm,bm}
\usepackage{mathtools}
\usepackage{slashed}
\usepackage{mathalfa}

\usepackage{graphicx,caption,subfigure}
\usepackage{tikz} 
\usepackage{tikz-cd}
\usetikzlibrary{decorations.pathreplacing,calc,tikzmark}
\usepackage{booktabs}
\usepackage{adjustbox}
\usepackage[utf8]{inputenc}
\usepackage[splitrule]{footmisc} 
 
\usepackage{placeins}
\usepackage{empheq}
\usepackage{paralist}
\usepackage{cite}
\usepackage[normalem]{ulem}
\usepackage{soul}

\usepackage[colorlinks=false,urlbordercolor=red
]{hyperref}
\usepackage{xcolor}
\usepackage{tensor}

\newcommand{\nc}{\newcommand}
\nc{\lb}{\llbracket}
\nc{\rb}{\rrbracket}
\nc{\gl}{\llbracket}
\nc{\gr}{\rrbracket}
\nc{\del}{\partial}
\nc{\eq}[1]{\begin{equation}
                     \begin{split} #1 \end{split}
                     \end{equation}
}
\nc{\ov}{\overline}
\nc{\fa}{\hat}
\nc{\fb}{\MakeUppercase}
\nc{\fc}{\tilde}
\nc{\myhash}{\raisebox{\depth}{\#}}

\allowdisplaybreaks[2]
\numberwithin{equation}{section}
\DeclareUnicodeCharacter{2212}{-}

\begin{document}


\vspace*{-1.5cm}
\begin{flushright}
  {\small
  MPP-2024-72
  }
\end{flushright}

\vspace{1.0cm}
\begin{center}
  {\huge  Emergence of $R^4$-terms in M-theory } 
\vspace{0.4cm}

\end{center}

\vspace{0.25cm}
\begin{center}
{
\Large Ralph Blumenhagen$^{1,2}$, Niccol\`o Cribiori$^1$, Aleksandar Gligovic$^{1,2}$, \\[.2cm] Antonia Paraskevopoulou$^{1,3}$
}
\end{center}

\vspace{0.0cm}
\begin{center} 
\emph{
$^{1}$ 
Max-Planck-Institut f\"ur Physik (Werner-Heisenberg-Institut), \\ 
Boltzmannstr.  8,  85748 Garching, Germany } 
\\[0.1cm] 
\vspace{0.25cm} 
\emph{$^{2}$ Exzellenzcluster ORIGINS, Boltzmannstr.~2, D-85748 Garching, Germany}\\[0.1cm]
\vspace{0.25cm} 
\emph{$^{3}$ Fakult{\"a}t f{\"u}r Physik, Ludwig-Maximilians-Universit{\"a}t M\"unchen, \\ 
Theresienstr.~37, 80333 M\"unchen, Germany}\\[0.1cm]
\vspace{0.3cm}
\end{center} 
\vspace{0.5cm}

\begin{abstract}
It has been recently suggested that the strong Emergence Proposal is realized in M-theory limits by integrating out all light towers of states with a typical mass scale not larger than the species scale, i.e. the eleventh dimensional Planck mass. 
Within the BPS sector, these are transverse $M2$- and $M5$-branes, that can be wrapped and particle-like, carrying Kaluza-Klein momentum along the compact directions. 
We provide additional evidence for this picture by revisiting and investigating further the computation of $R^4$-interactions in M-theory \`a la  Green-Gutperle-Vanhove. 
A central aspect is a novel UV-regularization of Schwinger-like integrals, whose actual meaning and power we clarify by first applying it to string perturbation theory.
We consider then toroidal compactifications of M-theory and provide evidence that integrating out all light towers of states via Schwinger-like integrals thus regularized yields the complete result for $R^4$-interactions. In particular, this includes terms that are tree-level, one-loop and space-time instanton corrections from the weakly coupled point of view.
Finally, we comment on the conceptual difference of our approach to earlier closely related work by Kiritsis-Pioline and Obers-Pioline, leading to a correspondence between two types of constrained Eisenstein series.
\end{abstract}

\thispagestyle{empty}
\clearpage

\setcounter{tocdepth}{2}

\tableofcontents


\section{Introduction}

The aim of the swampland program, see e.g.\cite{Palti:2019pca,Agmon:2022thq} for reviews, is to distinguish between those effective theories that admit an ultraviolet (UV) completion to quantum gravity and those that do not. This endeavor is currently based on an intricate web of conjectures mainly motivated by string theory and black hole physics. While some of them, such as the distance conjecture \cite{Ooguri:2006in} or the weak gravity conjecture \cite{Arkani-Hamed:2006emk}, are by now well-established and tested, this is not yet the case for the Emergence Proposal \cite{Palti:2019pca,Heidenreich:2017sim,Grimm:2018ohb,Heidenreich:2018kpg}.


In most of its formulations, the Emergence Proposal postulates that terms in the low-energy effective action are emerging after integrating out towers of states becoming exponentially light in asymptotic regions of the moduli space, in agreement with the distance conjecture. 
Going beyond the initially employed effective field theory approximation \cite{Palti:2019pca,Heidenreich:2017sim,Grimm:2018ohb,Heidenreich:2018kpg,Corvilain:2018lgw, Marchesano:2022axe, Castellano:2022bvr, Blumenhagen:2023yws}, in \cite{Blumenhagen:2023tev} evidence was provided for an exact realization of the Emergence Proposal in a specific decompactification limit within the vector multiplet moduli space of type IIA string theory preserving $\mathcal{N}=2$ supersymmetry in four dimensions.
Being at strong string coupling, this decompactification limit is the known M-theory lift, where a fifth spacetime direction grows large.

Concretely, the result of \cite{Blumenhagen:2023tev} has been achieved via the proposal of a specific regularization scheme for the naively UV-divergent Schwinger integrals as well as a physical principle dictating which states are to be considered perturbative and integrated out.
For the latter purpose, recent advances in the swampland program suggest that one needs to sum over the full infinite towers of states with typical mass scale not larger than the quantum gravity cut-off, which is taken to be the species scale \cite{Dvali:2007hz,Dvali:2007wp} (see e.g.~\cite{Veneziano:2001ah} for earlier work). This has been recently subject of intense investigation, especially in type IIA compactifications preserving eight supercharges, see e.g.\cite{vandeHeisteeg:2022btw,Cribiori:2022nke,vandeHeisteeg:2023ubh,vandeHeisteeg:2023uxj,Cribiori:2023ffn,Cribiori:2023sch,vandeHeisteeg:2023dlw,Castellano:2023aum}, where one can provide an explicit description also in the interior of the moduli space, including the so-called desert point \cite{Long:2021jlv}.

From the general analysis of \cite{Blumenhagen:2023xmk}, one can see that in
the M-theory limit of critical type IIA string theory the species scale is the eleven-dimensional Planck mass, while the light states are made of $D0$-branes, which are Kaluza-Klein modes longitudinal to the M-theory circle, together with transversal $M2$- and $M5$-branes.
When compactifying the setup on a Calabi-Yau threefold down to four dimensions, due to the special feature that the $\mathcal{N}=2$ prepotential is a 1/2-BPS saturated amplitude, a drastic simplification occurs and one can carry out an emergence computation without knowing the complete spectrum of M-theory, i.e.~in particular the modes coming from the excitations of $M2$- and $M5$-branes, or the spectrum of the BFSS matrix model \cite{Banks:1996vh} (see \cite{Bilal:1997fy,Bigatti:1997jy,Taylor:2001vb} for reviews), respectively.
Extending on   the seminal work of \cite{Gopakumar:1998ii,Gopakumar:1998jq},
in \cite{Blumenhagen:2023tev} a regularization of the divergent Schwinger
integrals was introduced which allowed 
to recover the non-ambiguous terms in  the type IIA prepotential and the genus
one topological amplitude of the resolved conifold,  in agreement with \cite{Gopakumar:1998ki}.
Note that this included the polynomial terms as well as the world-sheet instanton corrections.
Motivated by this, in \cite{Hattab:2023moj} a different method  to get the polynomial terms
in the prepotential was proposed, where the Schwinger integral was deformed
to a complex  contour integral  with the integrand  changed
such that the residues reproduce the sum over the exponential terms and also the polynomial
terms from the pole at the origin. The relation between the two approaches
is still unclear.

In \cite{Blumenhagen:2023tev} it was also observed that for an emergent (i.e.~perturbative) string limit over the same moduli space, the Schwinger integral could only provide the usual string one-loop amplitude. 
Given instead the positive outcome of the calculation performed in the strong type IIA string coupling limit, it has been suggested that the natural home for a possible realization of the Emergence Proposal is indeed M-theory (with large circle radius, $r_{11}\gg 1$), which as of today remains largely unknown. 
This resonates in a compelling way with the observation that  tree-level graviton interactions are induced by loop-diagrams in the BFSS matrix model \cite{Banks:1996vh}, which can be traced back to the familiar tree-channel loop-channel equivalence for the interaction of $D0$-branes \cite{Douglas:1996yp}. We are therefore led to an M-theoretic refinement of the Emergence Proposal.

\begin{quotation}
\noindent
{\it  Emergence Proposal (M-theory):
In the infinite distance M-theory limit, $M_* R_{11}\gg 1$, with the Planck scale kept fixed, a perturbative quantum gravity theory arises whose low energy effective description emerges via quantum effects by integrating out the full infinite towers of states with a mass scale parametrically not larger than the eleven-dimensional
Planck scale $M_*$. These are transverse $M2$-, $M5$-branes carrying momentum along the eleventh direction ($D0$-branes) and along any potentially present compact direction.}
\end{quotation}

The precise definition of the limit will be presented in section \ref{sec:rtothefour}.
It is the purpose of this work to further test the Emergence Proposal for M-theory with another interaction, which is  simple enough to admit  explicit and reliable Schwinger-like computations, while also rich enough to allow for more sophisticated tests. 
Concretely, we consider toroidal compactifications of M-theory and focus on the $R^4$-terms in the (type IIA) effective action.\footnote{$R^4$-couplings in low energy effective actions have been studied recently in relation to the species scale in \cite{Calderon-Infante:2023uhz, vandeHeisteeg:2023dlw, Castellano:2023aum}.}
Our analysis will be  built upon the approach  of Green-Gutperle-Vanhove (GGV) \cite{Green:1997as}, where this term was obtained via a four-graviton scattering at one-loop producing a Schwinger-like calculation in which bound states of $D0$-branes were integrated out (see also \cite{Russo:1997mk}, containing a discussion on cut-off dependent couplings in eleven dimensions). 
In fact, it was already noticed at that time that this $D0$-brane Schwinger integral in ten and nine dimensions reproduces correctly the tree-level and (Euclidean) $E\!D0$-brane instanton contributions to the $R^4$-term, whereas the one-loop correction could only be partially recovered from first principles. More precisely, an infinite constant had to be set to the correct finite value $2\pi^2/3$. Additionally, to obtain the full result in eight dimensions, string instantons needed to be added by hand.
Nevertheless, from our point of view the work of GGV is in accordance with a realization of the Emergence Proposal in M-theory, and the coefficient of the $R^4$-term is a well-suited testing ground for exploring and refining this idea further.

Similarly to the four-dimensional $\mathcal{N}=2$ topological amplitudes, also the $R^4$-term is strongly believed to be 1/2-BPS saturated \cite{Green:1997di}, so that we have a chance of proceeding without precise knowledge of the complete Hilbert space of M-theory. 
While its computation is sensitive to all 1/2-BPS states coupling to the graviton, as it happens in general for one-loop gravitational amplitudes, it is not restricted to four non-compact dimensions and, upon toroidal compactification, one can in principle perform the analysis in any $d\leq 10$. In this paper, we will focus on $d\ge 4$.
We will extend the GGV Schwinger-like integral for $D0$-branes to also include other particle-like 1/2-BPS states, resulting from wrapping the light objects, i.e.~transverse $M2$- and $M5$-branes, on appropriate $2$- and $5$-cycles of the internal torus.

While the expressions we obtain are inevitably very similar to those derived in the early work of \cite{Kiritsis:1997em,Pioline:1997pu,Obers:1998fb,Obers:1999um}, we can already anticipate two main differences. First, the guiding principle therein was U-duality of the final $R^4$-interactions; second, in their approach also longitudinal $M2$- and $M5$-branes were integrated out. 
In contrast, motivated by \cite{Blumenhagen:2023tev,Blumenhagen:2023xmk}, we work here in the asymptotic regime of large eleventh direction and as such we break part of the U-duality symmetry.
Besides, recovering the string one-loop correction is a non-trivial issue in our approach, especially given that we are not integrating out the fundamental string corresponding to the longitudinal $M2$-brane.

An essential step of the present work is again to employ a proper regularization scheme for the UV-divergent Schwinger integrals in a real variable $t$. 
We invoke the same method as in \cite{Blumenhagen:2023tev}, whose meaning we clarify by applying it to the regime of weak string coupling,  $g_s\ll 1$, i.e.~the emergent string limit. 
In this case, it is well-established that the real Schwinger parameter $t\equiv \tau_1$ is complexified to $\tau=\tau_0+i\tau_1$ by implementing the string level-matching condition, $L_0-\overline{L}_0=m_I n^I +N-\overline N=0$, via a Lagrange multiplier $\tau_0$ (introduced as a delta function) and using then modular invariance to restrict the complex integration to the fundamental domain to avoid the UV singularity. 
Notice that for vanishing string excitations, $N=\overline N=0$, level matching becomes the 1/2-BPS condition, $m_I \,n^I=0$, for Kaluza-Klein momentum $m_I$ and string winding modes $n^I$.
We will provide evidence that in spacetime dimensions $d\leq 9$ our regularization method indeed gives the same result as carrying out the fully-fledged integration over the modular parameter of the string world-sheet at genus one. 
Note that the recent work \cite{Hattab:2023moj} argues that Schwinger integrals for extended objects, such as wrapped $D2$-branes, should be probing the internal structure in the UV region, i.e.~close to the singularity at $t=0$.
In string theory, this is clearly happening. 
Since we arrive at the same result, we conclude that via implementing the 1/2-BPS conditions and performing the regularization as in the present work, we are also resolving this issue, at least implicitly.

From a more formal point of view, the objective of this paper is to analyze how far our strategy of evaluating Schwinger-like integrals can be generalized to the strong coupling/M-theory limit.
Here, the 1/2-BPS sector becomes more involved than that of the emergent string limit, and it is now given by bound states of transverse $M2$- and $M5$-branes carrying momentum along the eleventh direction.
The BPS conditions themselves are also more complicated and in particular are not given by just a single constraint.
In fact, the number of such conditions increases with the number of compact spacetime dimensions. 
This is reminiscent and in fact closely related to the so-called section conditions in Double Field Theory (DFT) and Exceptional Field Theory (ExFT), see e.g.~\cite{Berman:2020tqn} for a review.
Section conditions are in fact 1/2-BPS conditions, just written as differential operators on the extended space made from ordinary and winding coordinates.
Application of ExFT to the computation of four graviton amplitudes can be found in \cite{Bossard:2015foa}.

In contrast to string theory, in M-theory it is not yet known if there exists a higher dimensional version of Schwinger-like integrals in which the UV singularity is directly taken care of. 
Precisely for this reason, our method of regularizing such Schwinger-like integrals, which is rather bottom-up, becomes relevant and for the moment constitutes the only way to proceed and explore how the M-theory generalization might work.

Up to a few boundary cases to be mentioned, in all studied examples the Schwinger-like integrals seem to provide the complete result
for the coefficients of the $R^4$-term, i.e.~including those contributions that are tree-level, one-loop and space-time instanton corrections from the
weakly coupled point of view. In this respect, recovering the one-loop terms, which also contain all world-sheet instanton corrections, is highly non-trivial as the longitudinal $M2$-brane, i.e.~the fundamental string in type IIA, is not among the light degrees of freedom. Thus, our results  provide additional evidence for the Emergence Proposal in M-Theory, though still for a 1/2-BPS saturated coupling. 
Going beyond the BPS lamppost, namely considering non-supersymmetric couplings, is of clear importance, but it seems also out of reach at the moment. The analogy to the weakly coupled string case suggests that for general couplings, the 1/2-BPS conditions will get promoted to something that might be called ``M-theory level matching conditions'', in which also all the other excitations of M-theory, like excitations of $M2$- and $M5$-branes,  will contribute.

This paper is organized  as follows. In section \ref{sec2:preliminaryGGVst}, we present a calculation \`a la GGV for the emergent string limit of toroidal compactifications of type IIA string theory, while also clarifying the employed methods of \cite{Blumenhagen:2023tev} in a broader context. 
In section \ref{sec:rtothefour}, we present in detail the GGV computation in M-theory and perform the explicit emergence calculations for the $R^4$-terms in ten and nine dimensions. 
In section \ref{sec:rtothe48d7d}, we perform the same calculation in eight and seven dimensions and point out the implications for the nine- and ten-dimensional cases previously analyzed. 
In section \ref{sec:rtothe4d<7}, we discuss a couple of new aspects appearing for lower dimensional cases. 
Readers not interested in the derivation and motivation of our results are advised to take a look into section \ref{sec:exceptional}, where we provide a general view on our conceptual principles and results and put them into perspective to the early work \cite{Kiritsis:1997em,Pioline:1997pu,Green:1997di,Obers:1998fb,Obers:1999um}. 
Our swampland-motivated reasoning  finally guides us to a relation between two a priori different Eisenstein series, which we explicitly prove with our methods in dimensions $d\ge 8$, and which has been proven for $d\ge 4$ and more in general in \cite{Bossard:2015foa,Bossard:2016hgy}.
Our conclusions and outlook are summarized in section \ref{sec:conclusion}. 
Two appendices contain additional details on perturbative terms in seven dimensions and on linear Diophantine equations. 


\section{One-loop \texorpdfstring{$R^4$}{TEXT}-terms in perturbative string theory}
\label{sec2:preliminaryGGVst}

In this section we will provide some background for the calculation of the $R^4$-term \`a la GGV \cite{Green:1997as}. We will begin with the more familiar perturbative string theory limit, where the well-known one-loop contributions to the four-graviton amplitude will be recovered using the regularization technique of \cite{Blumenhagen:2023tev}. In doing so, we will briefly analyze its equivalence to that of GGV. The fact that at this limit we do not obtain the tree level term, as one would expect from the strong Emergence Proposal is also in agreement with the findings of \cite{Blumenhagen:2023tev} for an emergent string limit. The actual extension of the GGV formalism for the M-theory limit will be studied in detail in the following sections.

\subsection{Emergent string limit}
\label{string limit}

The main object of interest in the present work is the coefficient $a_d$ of the higher derivative $R^4$-term \cite{Green:1997di} 
\begin{equation}
\label{R4termdstring}
S_{R^4}\simeq M_s^{d-8}\int d^d x  \sqrt{-g}\,  \vartheta_{(k)}\, a_d\, t_8 t_8\, R^4
\end{equation}
appearing in the $d$-dimensional effective theory arising after compactifying type IIA string theory on a $k$-dimensional torus $T^k$.\footnote{Here and in the following, we will be neglecting overall constant normalizations, but we will be carefully keeping track of relative ones.} We denote the internal volume (in string frame) as $\vartheta_{(k)}\equiv \vartheta_{12\ldots k}=\rho_{1}\rho_{2}\ldots\rho_k$, where each index corresponds to a compact direction of radius $\rho_I$ in string units, while the metric is
in string frame.  
In order to keep the presentation simple, we restrict ourselves to rectangular tori and we set to zero a possible internal  NS-NS $B_2$-form. 
There exists already a plethora of work on the term \eqref{R4termdstring}. Our aim is to reconsider it from the viewpoint of the Emergence Proposal \cite{Heidenreich:2017sim,Grimm:2018ohb}, using it as an example highlighting the applicability of our regularization method.

Following \cite{Green:1997di}, we recall that after compactifying on $T^k$ the one-loop contribution to the four-graviton amplitude in $d=10-k$ dimensions is given by 
\begin{equation}\label{ad pert st general}
a_{10-k}^{\rm 1-loop}\simeq 2\pi \int_{\mathcal{F}}\frac{d^2\tau}{ \tau_1^2}\,\sum_{m^I,n^I\in\mathbb{Z}^k}e^{-\frac{\pi}{\tau_1}\sum_{I,J=1}^k (m^I+n^I\tau)G_{IJ}(m^J+n^J\bar{\tau})}\,,
\end{equation}
where $\tau=\tau_0+i\tau_1$ and the Gaussian integration over the continuous momenta along the $d$ non-compact directions has already been carried out. Here,  $G_{IJ}$ denotes  the torus metric and $\mathcal{F}$ the fundamental domain of $SL(2,\mathbb{Z})$ to which one restricts due to modular invariance. The vectors $m_I$ and $n^I$ respectively denote Kaluza-Klein momentum and winding along the $I$-th direction. As usual in string theory, these amplitudes are UV-finite due to removing the dangerous region close to the origin from the intergration region $\mathcal{F}$.

Undoing  the usual textbook derivation of such one-loop amplitudes, we can apply a Poisson resummation\footnote{
For a symmetric positive definite $k\times k$ matrix $G$ and a real
vector $b^I$, Poisson resummation amounts to $$\sum_{m^I\in
  \mathbb{Z}^k} e^{-\frac{\pi}{t}\sum_{I,J}(m^I+b^I)G_{IJ}(m^J+b^J)} =
\frac{t^{\frac k2}}{\sqrt{\det(G_{IJ})}}\sum_{m_I \in\mathbb{Z}^k}
e^{-2\pi i\sum_I m_I b^I - \pi t\sum_{I,J} m_I G^{IJ}m_J}.$$} of the Kaluza-Klein modes, extend the integration region to the full upper half plane and carry out the integral over $\tau_0$ to give the level-matching constraint, which for vanishing string excitation becomes the 1/2-BPS condition
\begin{equation}
\label{BPS strings}
\sum_{I=1}^{k} m_I \, n^I = 0\,.
\end{equation}   
In this way we can recast \eqref{ad pert st general} into the form
\begin{equation}
\label{adGGVstring}
a_{10-k}^{\rm {1-loop}} \simeq \frac{2\pi}{\vartheta_{(k)}}
\int_{0}^{\infty} \frac{dt}{t^\frac{4-k}{2}} \, \sum_{m_I, n^I \in
  \mathbb{Z}^k}\, \delta({\rm BPS}) \, e^{-\pi t M^2}\, ,
\end{equation}
with $t=\tau_1$ and the mass
\begin{equation}
M^2= m_I G^{IJ} m_J + n^I G_{IJ} n^J\,.
\end{equation}
Since now the integration goes down to $t=0$, this is a UV-divergent real Schwinger integral.
As we will see, in the M-theory setting one has to deal with a very similar expression, where however the number of 1/2-BPS conditions is larger than one and the stringy regularization \eqref{ad pert st general} is not applicable. Hence, the question is whether one can regularize the real Schwinger integral using a different method but still giving the same result as \eqref{ad pert st general}.
We will show in the sequel of this section that the regularization method that we employed in \cite{Blumenhagen:2023tev} does precisely fulfil this requirement.

To connect to previous work \cite{Obers:1999um}, let us observe that by redefining $t\rightarrow1/t$ and minimally subtracting the term $(m_I,n^I)=(0,0)$, the Schwinger integral  reduces to a constrained Eisenstein series\footnote{Following the conventions of \cite{Obers:1999um}, the Eisenstein series of order $s$ for a representation $\mathcal{R}$ of a group $\mathcal{G}$ is $$
\mathcal{E}^{\mathcal{G}}_{\mathcal{R};s}=\hat{\sum_{m_i\in\mathbb{Z}}}\left[\sum_{i,j}m_iM_{ij}m_j\right]^{-s}=\frac{\pi^s}{\Gamma(s)}\hat{\sum_{m_i\in\mathbb{Z}}}\int_0^\infty\frac{dt}{t^{s+1}}e^{-\frac{\pi}{t}\sum_{i,j}m_iM_{ij}m_j}\,,$$
where $M_{ij}$ transforms in $\mathcal{R}$ and
$\hat{\sum}_{m_i\in\mathbb{Z}}$ indicates that the term with all
$m_i=0$ has been excluded. The matrix $M^{ij}$ transforms in the
conjugate representation $\bar{\mathcal{R}}$.} of order $s=k/2-1$ in the vector
representation of $SO(k,k)$ (see also \cite{Angelantonj:2011br} for a more formal proof of this statement)
\begin{equation}\label{ad Eis int rep}
  \mathcal{E}^{SO(k,k)}_{V;s={\frac k2}-1}\simeq  \frac{2\pi}{\vartheta_{(k)}}\int_0^\infty\frac{dt}{t^{(k/2-1)+1}}\,\hat{\sum_{m_I, n^I \in
  \mathbb{Z}^k}}\delta({\rm BPS}) \,e^{-\frac{\pi}{t} M^2}\,.
\end{equation}
Such Eisenstein series are absolutely convergent only for $s>{\rm dim}(\mathcal{R})/2$. This is not the case for the string amplitude, which therefore is divergent and needs to be regularized, as usually done via analytic continuation in the $s$-plane. 
However, when dealing with these more general constrained Eisenstein series, such a step is non-trivial.
From this perspective, our physics motivated approach can be interpreted as a pragmatic recipe of how to perform the regularization at a microscopic level.

Let us mention that the expression \eqref{adGGVstring} can be considered a standard example of an emergence calculation. For an emergent string limit the species scale is identified with the string scale $M_s$, and light states are Kaluza-Klein and string winding modes. 
The 1/2-BPS constraint \eqref{BPS strings} is the familiar level matching condition for vanishing string oscillators. With respect to the $m_I$'s, this is a linear Diophantine equation in $k$ variables, which we are going to solve by employing the techniques of appendix \ref{app:Diophantine eqs}.

\subsection{Regularization methods}
\label{regularizations}

Let us explain the invoked regularization method from \cite{Blumenhagen:2023tev} in a very simple example, while commenting also on how certain seemingly different approaches to deal with UV divergences are, in fact, equivalent.
In the course of the computations, we will encounter Schwinger-like integrals of the type  
\begin{equation}
\label{start}
I= \sum_{m\in\mathbb Z} \int_0^\infty \frac{d t}{t^{s+1}} e^{-\pi t \frac{m^2}{r^2}}\,,
\end{equation}
with $s$ (half-)integer and $r>0$. 
For $s<0$ and $m\neq 0$, these integrals are convergent and can be computed directly, however the resulting series is convergent only for $s\neq-1/2$. For $s\geq 0$ and $m\neq 0$, they are divergent in the region $t\simeq 0$. The term $m=0$ requires yet further care.
To deal with this issue, in \cite{Green:1997as} the authors first applied a Poisson resummation with respect to the integer $m$, leading to 
\begin{equation}
I=\sum_{\hat m\in\mathbb Z}  r \int_0^\infty  d\hat t\; {\hat t}^{s-\frac12} \, e^{-\pi \hat t \,{\hat m^2  r^2}}  \,    .
\end{equation}
 with $\hat t=1/t$.
For $\hat m=0$ the integral is divergent, but for all other values of $\hat m$ (and for $s\geq 0$) it can be carried out explicitly to arrive at 
\begin{equation}
\label{regupoisson}
 I={\rm sing.} + 2\, \frac{\Gamma(s+1/2)}{ r^{2s} \, \pi^{s+1/2}} \sum_{\hat m\ge 1} \frac{1}{ {\hat m}^{2s+1}}={\rm sing.} + 2\, \frac{\Gamma(s+1/2)}{ r^{2s}  \,\pi^{s+1/2}}  \,  \zeta(2s+1)\,,
\end{equation}
where $\Gamma(s)$ is the $\Gamma$-function and $\zeta(s)=\sum_{n=1}^\infty n^{-s}$ is the Riemann $\zeta$-function.
Hence, in this way the singularity $(\hat m=0)$ has been isolated and can be subtracted. This method has also been used for the calculation of $R^4$- and $F^4$-terms in toroidal compactifications of type I/heterotic string theory in \cite{Bachas:1996bp,Bachas:1997mc}.

In \cite{Blumenhagen:2023tev}, we regularized such integrals in a manner that may seem to be different at first sight. 
There, the method was to introduce an ultraviolet cut-off $\epsilon$, with $\epsilon>0$, carry out the integral \eqref{start} directly and then expand the result around $\epsilon = 0\,$. 
For instance, for $s=1$ after these steps one has
\begin{equation}\label{int exp/t^2}
\int_{\epsilon}^{\infty}\frac{d t}{t^2}\,e^{-\pi t A}=\frac{1}{\epsilon}+\pi A \Big(\log(\pi A\epsilon) +\gamma_E -1\Big) +\mathcal{O}(\epsilon)\,,
\end{equation}
where $\gamma_E$ is the Euler-Mascheroni constant.
Then, one minimally subtracts those terms diverging in the region $\epsilon \simeq 0$ and eventually sends $\epsilon\to 0$ in the remaining expression.  
Proceeding in this way for non-negative, integer $s$ we get 
\begin{equation}
I= - \frac{(-1)^s \pi^s}{ r^{2s}\, s!} \sum_{m\in\mathbb Z} m^{2s} \log (m^2)  \,.
\end{equation}
The term $m=0$ now vanishes so that we can employ the derivative of $\zeta$-function to find
\begin{equation}
I= -\frac{4\,\pi^s (-1)^{s}}{ r^{2s} \,s!} \sum_{m\ge 1} m^{2s} \log m =  \frac{4\,\pi^s (-1)^s}{r^{2s}\, s!} \zeta'(-2s)\,.
\end{equation}
Using the two relations
\begin{equation}\label{zeta id1}
\zeta'(-2s)=\frac{(-1)^s(2s)!}{ 2 (2\pi)^{2s}} \zeta(2s+1)\,,\qquad \Gamma(s+1/2)=\frac{\pi^{1/2} (2s)!}{ 4^s \,s!}\,,
\end{equation}
one can see that this is the same as the non-singular term in
\eqref{regupoisson}. 
One can perform a similar computation for (positive) half-integer $s$. After minimal subtraction of divergent terms and then sending $\epsilon\to 0$, one gets
\begin{equation}
I= 2\frac{\Gamma(-s) \,\pi^s }{ r^{2s} } \sum_{m\ge 1 } m^{2s} =2\, \frac{\Gamma(-s)\, \pi^s }{ r^{2s} }  \zeta(-2s)\,.
\end{equation}
Now invoking the relation
\begin{equation}\label{gamma zeta id}
\zeta(-2s)=\frac{\Gamma(s+1/2)}{ \pi^{2s+{\frac 12}} \Gamma(-s)}\zeta(2s+1)\,,
\end{equation}
one obtains again the non-singular term in \eqref{regupoisson}.   

Thus,  we conclude that the methods of regularization of such Schwinger-like integrals used in \cite{Green:1997as} and in \cite{Blumenhagen:2023tev} are indeed equivalent. 
In the following, we will heavily apply our method of combining minimal substraction and $\zeta$-function regularization whenever we encounter diverging integrals of this type.

\subsection{One-loop corrections to \texorpdfstring{$R^4$}{TEXT}-terms in \texorpdfstring{$d\leq10$}{TEXT} dimensions}
\label{sec:R4stringth}

We now start employing this method to evaluate one-loop real Schwinger integral \eqref{adGGVstring} for different values of $k$.
Similarly to what is done with the Eisenstein series \eqref{ad Eis int rep}, we will also remove the contribution  $(m_I,n^I)=(0,0)$ from  the Schwinger integral. 
Comparing the result with the known string theory amplitude \eqref{ad pert st general}, we will see that our straightforward regularization method produces meaningful results.

\subsubsection*{One-loop \texorpdfstring{$R^4$}{TEXT}-term in ten dimensions}
In ten dimensions, the string theory formula \eqref{ad pert st general} gives the known expression
\begin{equation}
\label{a10st1loop}
a_{10}^{\rm 1-loop}\simeq 2\pi \int_\mathcal{{F}}\frac{d^2\tau}{\tau_1^2}=\frac{2\pi^2}{3}\,.
\end{equation}
On the other hand, when specializing \eqref{adGGVstring} to $d=10$ ($k=0$) and employing the regularization scheme introduced in the previous section, we get a vanishing result. Indeed, we find $\int_\epsilon^\infty dt / t^{2} = 1/\epsilon$, which is then minimally subtracted.
However, we will see below how the correct non-vanishing constant \eqref{a10st1loop} can be reproduced from \eqref{adGGVstring} after taking the decompactification limit of the regularized expression obtained for $d<10$. This will constitute a non-trivial consistency check for our approach.

\subsubsection*{One-loop \texorpdfstring{$R^4$}{TEXT}-term in nine dimensions}

In nine dimensions, namely for type IIA string theory compactified on $S^1$, the coefficient of the $R^4$-term at one-loop in string perturbation theory reads \cite{Green:1997di}
\begin{equation}
\label{a9st1loop}
a_{9}^{\rm 1-loop} \simeq \frac{2\pi^2}{3} \left( 1 + \frac{1}{\rho_{1}^2} \right)\,,
\end{equation}
with $\rho_{1}=R_{1}M_s$ the radius of the circle in string units.
To reproduce \eqref{a9st1loop}, we specialize \eqref{adGGVstring} to $d=9$ $(k=1)$. From the BPS condition, $m \cdot n = 0$, we see that two different sectors contribute, namely one with only winding, $n \neq 0$, and one with only Kaluza-Klein momentum, $m \neq 0$. As explained in section \ref{regularizations}, after subtracting the $m=n=0$ term, we proceed as in  \cite{Blumenhagen:2023tev}: we first introduce a UV regulator $\epsilon>0$ to calculate the integral and then carry out the sum over $m$ employing $\zeta$-function regularization.
Hence, in the first step we expand around $\epsilon\simeq 0$ to get
\begin{equation}\label{int exp/t^(3/2)}
\int_{\epsilon}^{\infty}\frac{d t}{t^{\frac 32}}e^{-t A}=\frac{2}{\sqrt\epsilon}-2\sqrt{\pi\,A}+\mathcal{O}(\sqrt{\epsilon})\,.
\end{equation}
Performing then the infinite sum, the winding sector contributes as
\begin{equation}
\label{string9dcontribution}
a_{9,m=0}^{{\rm 1-loop}} \simeq \frac{2\pi}{\rho_{1}} \sum_{n \neq 0} \int_{0}^{\infty} \frac{dt}{t^{3/2}} e^{-\pi t \rho_{1}^2 n^2} = \frac{2\pi^2}{3}\,,
\end{equation}
where we also used that $\zeta(-1)=-1/12$. 
Similarly, for the Kaluza-Klein sector one obtains
\begin{equation}
\label{a9n=0st}
a_{9,n=0}^{\rm 1-loop} \simeq \frac{2\pi}{\rho_{1}} \sum_{m \neq 0} \int_{0}^{\infty} \frac{dt}{t^{3/2}} e^{-\pi t\frac{ m^2}{\rho_{1}^2}} = \frac{2\pi^2}{3} \frac{1}{\rho_{1}^2}\,.
\end{equation}
Adding up these two expressions, the full one-loop result \eqref{a9st1loop} is recovered. Besides, the decompactification limit $\rho_1 \to \infty$ reproduces precisely the constant \eqref{a10st1loop}. From an M-theory point of view, the nine-dimensional result has also been recoved via a Schwinger integral integrating out the longitudinal $M2$-brane, i.e. the type IIA fundamental string, and Kaluza-Klein contributions in \cite{deWit:1999ir}.

\subsubsection*{One-loop \texorpdfstring{$R^4$}{TEXT}-term in eight dimensions}

In eight dimensions, namely for type IIA string theory compactified on $T^2$, the coefficient of the $R^4$-term at one-loop in string perturbation theory reads \cite{Green:1997as}
\begin{equation}
\label{a8d1loop string un}
a_8^{\rm 1 -loop} \simeq -\frac{2\pi}{\vartheta_{12}} \log\Big(\rho_2^2 |\eta(iu)\eta(i\vartheta_{12})|^4\Big) ,
\end{equation}
where the 2-torus volume is given by $\vartheta_{12}=\rho_1\rho_2$ and we introduced the complex structure modulus $u = \rho_2/\rho_{1}$, while $\eta(x)$ is the Dedekind $\eta$-function. Recall that we are working with an orthogonal torus for simplicity. 

To reproduce \eqref{a8d1loop string un}, we specialize \eqref{adGGVstring} to $d=8$ $(k=2)$.
The 1/2-BPS constraint \eqref{BPS strings} reads $m_1 n_1 + m_2 n_2 = 0$ and its solution leads to the identification of three sectors with different parametrizations of $(n_1,n_2,m_1,m_2)$, from which we will be excluding the $(0,0,0,0)$ contribution:\footnote{Not to confuse $n^I$ with the $I$-th power of $n$, in the following (with the exception of section \ref{sec:exceptional}) we write winding numbers with indices downstairs, $n^I\to n_I$. Momenta will always be denoted as $m_I$, so that one can distinguish them from winding.}
\begin{enumerate}
\item $n_1 = 0=n_2$. These states have unrestricted Kaluza-Klein momentum and no winding. We paremetrize them as $(0,0,m_1,m_2)$ with $m_1,m_2 \in \mathbb{Z}$.
\item $n_1 \neq 0$ or $n_2\neq 0$. These states have one unrestricted Kaluza-Klein momentum. We parametrize them as $(n_{1},0,0,m_{2})$ or $(0,n_{2},m_{1},0)$ with $n_{1,2},m_{1,2} \in \mathbb{Z}$.
\item $n_1 \neq 0 $ and $n_2 \neq 0$. In this case, the BPS condition becomes a linear homogeneous Diophantine equation in $m_1$ and $m_2$, whose solution is reviewed in appendix \ref{app:Diophantine eqs} and reads $m_1=M \tilde n_2$, $m_2=-M\tilde n_1$ with $M\in \mathbb{Z}$ and $\tilde n_{1,2} = n_{1,2}/{\rm gcd}(n_1,n_2)$. Since multiplication by an overall integer maps solutions into solutions, we can parametrize these states as $(N  \tilde n_{1}, N  \tilde n_{2}, M \tilde n_2,-M \tilde n_1)$, with $N \neq 0$ and $M \in \mathbb{Z}$.
\end{enumerate}
We now solve the Schwinger integral \eqref{adGGVstring} for these three cases separately and eventually combine the results. 
In the first case, we are dealing with the expression
\begin{equation}
a_8^{{\rm 1-loop},(1)} \simeq \frac{2\pi}{\vartheta_{12}} \sum_{(m_1,m_2)\neq(0,0)} \int_{0}^{\infty} \frac{dt}{t} e^{-\pi t \left( \frac{m_1^2}{\rho_{1}^2} + \frac{m_2^2}{\rho_{2}^2} \right)} \,.
\end{equation}
To calculate it, we split the sum over Kaluza-Klein momenta according to $\sum_{(m_1,m_2)\neq(0,0)}=\sum_{m_1\neq0,m_2\in\mathbb{Z}}+\sum_{m_1=0,m_2\neq0}$. We start from the first term, i.e.~the one with $m_1 \neq 0$, and perform a Poisson resummation over $m_2$ to obtain
\begin{equation}
a_{8,m_1 \neq 0}^{{\rm 1- loop},(1)} \simeq \frac{2\pi}{\rho_{1}} \sum_{m_1 \neq 0} \sum_{m_2 \in \mathbb{Z}} \int_{0}^{\infty} \frac{dt}{t^{3/2}} e^{ - \pi t \frac{m_1^2}{\rho_{1}^2} -\frac{\pi}{t} m_2^2 \rho_2^2 }.
\end{equation}
While the contribution $m_2=0$ can be computed as in the nine-dimensional case, see \eqref{a9n=0st}, the sum with $m_2 \neq 0$ is conveniently evaluated via the integral representation \cite{Kiritsis:1997em}
\begin{equation}
\label{besselrel}
\int_0^\infty \frac{dx}{x^{1-\nu}} \,e^{-{\frac{b}{x}}-cx}=2 \left|  {\frac{b}{c}}\right|^{\frac{\nu}{2}} K_\nu\Big(2\sqrt{|b\, c|}\Big)\, ,
\end{equation}
where $K_\nu(x)$ denotes the modified Bessel function of order $\nu$. In this way, using also that $K_{-\nu}(x)=K_\nu(x)$ and $K_{1/2}(x)=\sqrt{\frac{\pi}{2x}} e^{-x}$, we get
\begin{equation}
\begin{aligned}
a_{8,m_1\neq 0}^{{\rm 1 -loop},(1)} \simeq \frac{2\pi}{\vartheta_{12}} \left( \frac{\pi}{ 3} u + 4\!\! \sum_{m_1,m_2\geq 1} \frac{1}{m_2} e^{-2\pi m_1 m_2 u} \right) = - \frac{2\pi}{\vartheta_{12}} \log\left(|\eta(iu)|^4 \right)\,.
\end{aligned}
\end{equation}
Next we apply our regularization scheme to the remaining term $m_1=0$, using 
\begin{equation}\label{int exp/t}
\int_\epsilon^\infty\frac{d t}{t}e^{-t A}=-\gamma_E-\log(\epsilon A)+\mathcal{O}(\epsilon)\,.
\end{equation}
After minimally subtracting $\log(\epsilon)$,  this implies that
\begin{equation}
a_{8,m_1=0}^{{\rm 1-loop},(1)}\simeq \frac{2\pi}{\vartheta_{12}}\sum_{m_2\neq 0}\int_0^\infty\frac{dt}{t}e^{-\pi t\frac{m_2^2}{\rho_2^2}}\simeq\frac{2\pi}{\vartheta_{12}}\sum_{m_2\neq 0}\log(\rho_2^2)\simeq-\frac{2\pi}{\vartheta_{12}}\log(\rho_2^2)\,,
\end{equation}
where we have chosen a regulator $\tilde\epsilon=\epsilon\, 4\pi e^{-\gamma_E}$, compatible with modular invariance.\footnote{Any ambiguity such as dependence on the regulator scale reflects that our expressions correspond to an Eisenstein series of order $s=\rm{dim}(\mathcal{R})$, exhibiting a pole obstructing analytic continuation. In these cases, additional arguments, like e.g.~modular invariance, are used to choose the appropriate scale (as was also observed in \cite{Green:2010sp}).  When analytic continuation is possible, our regularization method does not exhibit such ambiguities.} Therefore, the contribution of the first set of states reads
\begin{equation}
\label{a81loopst1}
a_{8}^{{\rm 1-loop},(1)}\simeq - \frac{2\pi}{\vartheta_{12}}\log \left(\rho_2^2 |\eta(iu)|^4\right).
\end{equation}

There is a convenient way to perform the calculation simultaneously over the set of states $2$ and $3$ identified above. 
Let us denote by $\tilde{\mathbf{n}}_\alpha$ a $k$-dimensional array of non-vanishing coprime integers, such that the vector $\alpha_i\in \{0,1\}$ indicates which numbers are non-vanishing. 
The total number is then denoted as $|\alpha|=\sum_{i=1}^k \alpha_i$.
Hence $\alpha=(1,0)$ and  $\alpha=(0,1)$ correspond to case $2$ with $n_1\neq 0$ or $n_2 \neq 0$, respectively.
Similarly, $\alpha=(1,1)$ corresponds to  case $3$ with both  $n_1 \neq 0$ and $n_2\neq 0$. 
Since \eqref{adGGVstring} is invariant under $n^I \to - n^I$, we can restrict the winding sum to run over strictly positive integers only, at the cost of multiplying the integral by an overall factor $2^{|\alpha|}$. 
We can thus write the expression
\begin{equation}
\label{a81loopst23}
a_{8}^{{\rm 1 -loop},\alpha} \simeq \frac{2\pi }{\vartheta_{12}} \cdot 2^{|\alpha|}\sum_{\tilde{ \mathbf n}_\alpha >0} \sum_{M \in \mathbb{Z}} \sum_{N > 0} \int_{0}^{\infty} \frac{dt}{t}\; e^{-\pi t \left( N^2 L_\alpha^2+  \frac{M^2 L_\alpha^2}{\vartheta_{12}^2} \right)}\,,
\end{equation}
where $L_\alpha = \sqrt{{\tilde n}_1^2\rho_1^2+{\tilde
    n_2^2}\rho_2^2}$ and $\sum_{\tilde {\mathbf n}_\alpha}$ is
shorthand for a summation over the coprime integers described by $\alpha$.  For
$|\alpha|=1$, when say $n_1 \neq 0$ we set $\tilde n_2=0$ and $\tilde
n_1=1$, being the only solution to gcd$(\tilde n_1)$=1.
With these rules, the integral \eqref{a81loopst23} covers both the set
of states 2 and 3 at the same time.

Next, we perform a Poisson resummation over $M$ yielding
\begin{equation}
a_{8}^{{\rm 1 -loop},\alpha} \simeq 2\pi \cdot 2^{|\alpha|} \sum_{{\tilde{\mathbf n}}_\alpha >0}\frac{1}{L_\alpha} \sum_{M \in \mathbb{Z}} \sum_{N>0} \int_0^\infty \frac{dt}{t^{\frac 32}}\;e^{-\pi t N^2 L_\alpha^2 -\frac{\pi}{t}\frac{M^2\vartheta_{12}^2}{L_\alpha^2}}\,.
\end{equation}
Then, we split the sum into the contributions $M=0$ and $M\neq 0$. In the first case, after $\zeta$-function regularization for the infinite sum over $N$, we find a constant term
\begin{equation}
a_{8,M=0}^{{\rm 1 -loop},\alpha} \simeq \frac{2\pi^2}{3}2^{|\alpha|-1}\sum_{\tilde{\mathbf n}_\alpha>0}1\,.
\end{equation}
In the second case, we apply \eqref{besselrel} and the two different winding contributions give 
\begin{equation}
a_{8,M\neq 0}^{{\rm 1 -loop},\alpha}\simeq \frac{8\pi}{\vartheta_{12}} 2^{|\alpha|-1} \left(\sum_{\tilde{\mathbf n}_\alpha>0}1\right)\sum_{M>0}\sum_{N>0}\frac{1}{M}e^{-2\pi MN \vartheta_{12}}\,.
\end{equation}

Note that in both expressions $L_\alpha$ cancelled out so that they  only depend on $|\alpha|$.
Combining them, the integral \eqref{a81loopst23} results in
\eq{
\label{a81loopsta}
a_{8}^{{\rm 1 -loop},\alpha} &\simeq \frac{2\pi }{\vartheta_{12}} 2^{|\alpha|-1}\left(\frac{\pi}{3}\vartheta_{12}+4\sum_{M>0}\sum_{N>0}\frac{1}{M}e^{-2\pi MN \vartheta_{12}}\right)\sum_{\tilde{\mathbf n}_\alpha>0}1\\
&\simeq- \frac{2\pi}{\vartheta_{12}} 2^{|\alpha|-1}\log\left(|\eta(i\vartheta_{12}|^4\right)\sum_{\tilde{\mathbf n}_\alpha>0}1 \,.
}
We now have to regularize the sum over the array of coprime positive
integers $\tilde{\mathbf{n}}_\alpha$. For  this purpose, we invoke the relation 
\begin{equation}
\label{regcoprime}
\left(\sum_{\tilde{\mathbf n}_\alpha>0}1\right)\left(\sum_{N>0}1\right) = \sum_{{\mathbf n}_\alpha>0}1 \equiv \zeta(0)^{|\alpha|}\,,
\end{equation}
where we recall that $\tilde{\mathbf n}_\alpha$ are coprime integers while ${\mathbf n}_\alpha$ are just integers. We thus have that 
\begin{equation}
\label{regcoprime2}
\sum_{\tilde{\mathbf n}_\alpha>0} 1\equiv \zeta(0)^{|\alpha|-1} = \left(-\frac12\right)^{|\alpha|-1}
\end{equation}
and \eqref{a81loopsta} becomes 
\begin{equation}
a_{8}^{{\rm 1 -loop},\alpha} \simeq (-1)^{|\alpha|} \frac{2\pi }{\vartheta_{12}}\log \left(|\eta(i\vartheta_{12})|^4\right).
\end{equation}
Remarkably, the parameter $\alpha$ controls only an overall sign.
The contribution of the set of states $2$ and $3$ is now given by the
sum over $\alpha=(1,0),(0,1),(1,1)$ which we can write as\footnote{We used
\begin{equation}
\label{sumbinomial}
\sum_{l=0}^k  \binom{k}{l}(-1)^{l} =0,
\end{equation}
following from the Binomial Theorem $(x+1)^k = \sum_{l=0}^k\binom{k}{l}x^l$ evaluated at $x=-1$.}
\begin{equation}
\begin{aligned}
\label{a81loopst23fin}
a_{8}^{{\rm 1 -loop},(2)+(3)} &= \sum_{|\alpha|=1}^2 \binom{2}{|\alpha|}\,a_{8}^{{\rm 1 -loop},\alpha}= -\frac{2\pi }{\vartheta_{12}}\log \left(|\eta(i\vartheta_{12})|^4\right)\,.
\end{aligned}
\end{equation}
Adding up \eqref{a81loopst1} and \eqref{a81loopst23fin} we recover the
known expression \eqref{a8d1loop string un}. Clearly, in this
computation
$\zeta$-function regularization played an essential role.

\subsubsection*{One-loop \texorpdfstring{$R^4$}{TEXT}-terms in \texorpdfstring{$d\leq 7$}{TEXT}}

In $d\leq 7$ dimensions the situation becomes richer since solving the BPS condition \eqref{BPS strings} requires solving a linear Diophantine equation in $k \geq 3$ variables $m_I$, with the winding numbers $n_I$ as coefficients. Below we focus on the constant term, $2\pi^2/3$, and show that it remains the same for any $d\leq 7$, while the perturbative (in $g_s$) world-sheet instanton corrections will be examined in more detail in appendix \ref{7d pert terms} for the case $d=7$.

As reviewed in appendix \ref{app:Diophantine eqs}, solving a linear
Diophantine equation in $k$ variables requires the introduction of
$k-1$ integers for the Kaluza-Klein momenta $m_I$. This corresponds to
the situation in which all $k$ winding numbers are
non-vanishing. Setting any winding number to zero leads to the
corresponding Kaluza-Klein momentum being unconstrained, so that we
end up once more with $k-2+1 = k-1$ integers. This observation holds
for any number of vanishing $n_I$ as described by a vector $\alpha$, the only exception being the case in which all of them are vanishing as there is no Diophantine equation to be solved anymore. Physically, this corresponds to the pure Kaluza-Klein sector of the computation. We will not consider this sector in what follows, but rather assume that at least one of the winding numbers is non-vanishing. Hence, we will show that the constant contribution, $2\pi^2/3$, is related to extended objects, in this case strings, a point which we will return to in section \ref{sec:decompact}.

We have thus motivated the fact that in $d=10-k$ dimensions the momentum part of the Schwinger integral, excluding the pure Kaluza-Klein sector, can be parametrized by $k-1=9-d$ unconstrained integers $\mu_i$ appearing in the quadratic form
\begin{equation}
Q_\alpha^2=\mu_i\, {\cal M}_\alpha^{ij}\, \mu_j\,,
\end{equation} 
where the components of the $\alpha$-dependent momentum matrix ${\cal M}_\alpha^{ij}$ can be read off by re-placing $m_I$ in $\sum_{I=1}^k m_I^2/\rho_I^2$ with the solution of the appropriate linear Diophantine equation, i.e.~the BPS condition, and rewriting the resulting expression as a quadratic form in the integers $\mu_i$.\footnote{E.g.~for $k=4$ the integers $\mu_i$ correspond to $L,M,N$ in \eqref{soldioph4v1}-\eqref{soldioph4v2}. For $d=5$ the explicit expression of ${\cal M}^{ij}$ is reported in \eqref{MMatrix5}.}
The contribution of a configuration $\alpha$  with at least one
non-vanishing winding number
can be compactly written as 
\begin{equation}
\label{a81loopstw}
a_{d}^{\rm 1-loop,\alpha}\simeq\frac{2\pi}{\vartheta_{(k)}}\,2^{|\alpha|} \sum_{\tilde{\mathbf n}_\alpha>0}\sum_{\mu_i\in\mathbb{Z}}\,\sum_{N>0}\int_0^\infty\frac{dt}{t^\frac{4-k}{2}}\,
e^{-\pi  t\left(N^2L_\alpha^2+\mu_i {\cal M}_\alpha^{ij}\mu_j\right)}\,,
\end{equation}
with $L_\alpha^2=\sum_{I=1}^k \tilde n_I^2\rho_I^2$. Since the momentum matrix ${\cal M}_\alpha^{ij}$ is invariant under sign flips, provided the signs of the particular solutions of the corresponding Diophantine equations are flipped accordingly, it is possible to restrict the sums to positive winding numbers, giving rise to a symmetry factor $2^{|\alpha|}$. Besides, notice that the momentum matrix ${\cal M}_\alpha^{ij}$ satisfies ${\rm det}({\cal M}_\alpha^{ij})=L_\alpha^2/\vartheta_{(k)}^2$. We checked this explicitly for $d\geq 2$ (see appendix \ref{app:Diophantine eqs} for $d=5$) and believe it holds in any dimension. Poisson resummation over the $k-1$ integers $\mu_i$ produces a polynomial term $t^{\frac{1-k}{2}}$ which combines with the factor $t^{\frac{k-4}{2}}$ already present in \eqref{a81loopstw} resulting in $t^{-\frac32}$. Hence, after Poisson resummation we have
\begin{equation}
a_{d}^{\rm 1-loop,\alpha}\simeq 2\pi \,\sum_{\tilde{\mathbf n}_\alpha>0}\frac{2^{|\alpha|}}{L_\alpha}\sum_{\mu^i\in\mathbb{Z}^{k-1}}\, \sum_{N>0}\int_0^\infty\frac{dt}{t^\frac{3}{2}}\, e^{-\pi  t N^2L_\alpha^2-\frac{\pi}{t} \mu^i {\cal M}_{\alpha,ij}\mu^j}\,.
\end{equation}
The constant term arises now from the sector $\mu^i=0$ (for
$i=1\ldots,k-1$) in which case $L_\alpha$ cancels and, as for $d=8$,
the contribution only depends on $|\alpha|$.
Summing over all $\alpha$ and using $\zeta$-function regularization for the infinite sum over $N$, we find
\begin{equation}\label{ad wind const general}
a_{d,{\mu}^i=0}^{\rm 1-loop}  \simeq -4\pi^2 \zeta(-1) \sum_{|\alpha|=1}^k\,2^{|\alpha|} \binom{k}{|\alpha|}\sum_{\tilde{\mathbf n}_\alpha>0}1 = \frac{2\pi^2}{3}\sum_{|\alpha|=1}^k(-1)^{|\alpha|-1}\binom{k}{|\alpha|} = \frac{2\pi^2}{3}\,,
\end{equation}
where we used first \eqref{regcoprime2} and then \eqref{sumbinomial}. This proves that the constant term is the same in any dimension and originates from the contribution of extended objects, which in this limit are strings winding around the compact directions. 

The various sectors with $\vec\mu\neq\vec{0}$ give rise to world-sheet instanton contributions which, after using \eqref{besselrel}, can be written as 
\begin{equation}\label{ad wind instanton general}
\begin{aligned}
a_{d,{\vec\mu}\neq \vec 0}^{\rm 1-loop,\alpha}&\simeq 2\pi \,\sum_{\tilde{\mathbf n}_\alpha>0}\sum_{\vec{\mathbf{\mu}}\neq \vec{0}}\sum_{N>0}\frac{2^{|\alpha|}}{L_\alpha\hat  Q_\alpha}\,e^{-2\pi NL_\alpha \hat Q_\alpha}\,,
\end{aligned}
\end{equation}
with $\hat Q_\alpha^2=\mu^i {\cal M}_{\alpha,ij}\mu^j$.
In appendix \ref{7d pert terms} we present in detail the evaluation of all these terms for the specific case $d=7$ ($k=3$).
Again employing $\zeta$-function regularization we find that the final result agrees with the evaluation of the corresponding string amplitude \eqref{ad pert st general}, as e.g.~reported in \cite{Kiritsis:1997em}, eqs.~(5.16), (5.18), (5.20), except for the overall normalization of the triple instanton term.

To summarize, in  this section we have collected ample evidence that the proposed regularization for the real Schwinger integrals, via minimal substraction and $\zeta$-function regularization, gives correct results.
In lower non-compact dimensions, the actual computation will become more and more cumbersome, but is in principle a straightforward exercise.


\section{Emergence of \texorpdfstring{$R^4$}{TEXT}-terms in M-theory in \texorpdfstring{$d=10,9$}{TEXT}}
\label{sec:rtothefour}

In this section, we calculate the $R^4$-terms in toroidal compactifications of M-theory to ten and nine dimensions by evaluating the appropriate real Schwinger integrals, employing the same methods developed in the previous section. 
These are the simplest cases where light particle-like contributions can come only from Kaluza-Klein momenta (including $D0$-branes). Hence, our computation is inevitably very similar to the original work of GGV \cite{Green:1997as} but it also differs from it at some essential places that we are going to point out, setting the stage for its extension in the following sections. Before proceeding with these calculations we review the M-theory framework and the GGV approach in that limit.

\subsection{M-theory limit}

We begin by recalling the dictionary relating M-theory to the strong coupling limit of the type IIA superstring.
The string scale $M_s$ and coupling $g_s$ are given in terms of the eleven-dimensional Planck scale $M_*$ and of the size $R_{11}$ of the eleventh direction as
\begin{equation}
M_s^2=M_*^3\, R_{11}\,,\qquad g_s=(M_*\, R_{11})^{\frac{3}{2}}\,.
\end{equation}
We compactify the type IIA string on an internal space $X$ of dimension $k$ and consider a strong coupling limit, $\lambda\to\infty$, such that the $(d=10-k)$-dimensional Planck scale $M_{\rm pl}^{(d)}$ and the size of the internal space remain finite in units of $M_*$. In terms of M-theory quantities, this limit is given by
\begin{equation}
R_{11}\to \lambda  R_{11}\,,\qquad M_*\to \frac{M_*}{ \lambda^{\frac{1}{d-1}}}\,,\qquad R_I\to \lambda^{\frac{1}{d-1}} R_I,
\end{equation}
where $R_I$ denotes the physical radius of the internal space with volume ${V}_k=R_I^k$.
Indeed, using
\begin{equation}
\left(M_{\rm pl}^{(d)}\right)^{d-2}= M_*^9\, {V}_k\, R_{11},
\end{equation}
one can check that the $d$-dimensional Planck scale stays finite. 
Similarly, employing 
\begin{equation}
\left(M_{\rm pl}^{(d+1)}\right)^{d-1}= M_*^9\, {V}_k \,,
\end{equation}
the $(d+1)$-dimensional Planck scale $M_{\rm pl}^{(d+1)}$
scales in the same way as $M_*$. 
In terms of the type IIA quantities, this limit reads
\begin{equation}
g_s\to \lambda^{\frac{3(d-2)}{2(d-1)}}   g_s\,,\qquad M_s\to \lambda^{\frac{d-4}{2(d-1)}} \,M_s\,,\qquad R_I\to \lambda^{\frac{1}{d-1}} R_I\,.
\end{equation}  
Note that $d=4$ is special in the sense that the string scale
does not scale with $\lambda$.

As is well-known, from the M-theory perspective this particular type IIA strong coupling limit corresponds to decompactification from $d$ to $d+1$ dimensions. 
The lightest tower of states are $D0$-branes, or equivalently Kaluza-Klein states of the eleventh direction, and the resulting species scale $\tilde\Lambda$ is the $(d+1)$-dimensional Planck mass. As discussed in \cite{Blumenhagen:2023xmk}, transversally (to $S^1$) wrapped $M2$- and $M5$- branes and Kaluza-Klein modes along the compact space $X$ lead to a typical mass scale
\begin{equation}
\tilde\Lambda\sim M_{\rm pl}^{(d+1)}\sim M_*\,.
\end{equation}

The main object of interest is the coefficient $a_d$ of the higher derivative $R^4$-term \cite{Green:1997di}, 
\begin{equation}
\label{R4termd}
S_{R^4}\simeq M_*^{d-8}\int d^d x  \sqrt{-g}\, \mathcal{V}_k r_{11}\, a_d\, t_8 t_8\, R^4,
\end{equation}
in the $d$-dimensional effective theory arising after compactifying M-theory on a $(k+1)$-dimensional torus with volume $\mathcal{V}_k \, r_{11}$ (in units of $M_*$).
Here, the metric is still in string frame.  
Again, we restrict ourselves to rectangular tori and $B_2=0$. Our aim is to review and extend the computation from the viewpoint of the Emergence Proposal in M-theory \cite{Blumenhagen:2023tev,Blumenhagen:2023xmk}. 
The latter suggests that the term \eqref{R4termd} should emerge from a Schwinger-type one-loop integral over the aforementioned light towers of states, whose mass scale lies not higher than the species scale. 
Similarly to the four-dimensional setup studied recently in \cite{Blumenhagen:2023tev}, here we are also dealing with a 1/2-BPS saturated amplitude, a fact which makes the computation feasible since we can generically trust BPS masses also at strong coupling.
 
As in the emergent string limit, the computation of this term in M-theory, performed originally in the work by GGV \cite{Green:1997as}, is already very much in the spirit of the Emergence Proposal. 
Namely, they proposed that the complete coefficient $a_d$ can be computed  by compactifying M-theory on a $(k+1)$-dimensional torus with volume ${\cal V}_{k}r_{11}$ and by integrating out the Kaluza-Klein modes at one-loop.
They expressed the corresponding contributions as a Schwinger-like integral
\begin{equation}
\label{anGGV}
a_{10-k}\simeq \frac{1}{r_{11}{\cal V}_{k}} \int d^{10-k} \mathbf p \int_0^\infty {\frac{d t}{t}} t^4 \sum_{m_I\in \mathbb Z} e^{-\pi t\left( \mathbf p^2 + \sum_{I,J=1}^km_I  G_{(k+1)}^{IJ} m_J\right)}\,,
\end{equation}
where the factor $t^4$ arises from considering a four-graviton amplitude, while $G_{{(k+1)}\,IJ}$ is the metric of the $(k+1)$-torus. Moreover, all masses and volumes in \eqref{anGGV} are
measured in eleven-dimensional units, i.e.~in units of $M_*$.
Carrying out the integral over the  continuous  momenta $ \mathbf p$ gives
\begin{equation}
\label{d0emerge}
 a_{10-k}\simeq \frac{2\pi}{r_{11}{\cal V}_{k}}   \sum_{m_I\in\mathbb Z} \int_0^\infty \frac{dt}{t^{\frac{4-k}{2}}}  e^{-\pi t\, \sum_{I,J=1}^k m_I G_{(k+1)}^{IJ}  m_J}\,.
\end{equation}
In this expression, one integrates out the tower of $D0$-branes as well as Kaluza-Klein momenta along the internal $T^{k}$ (times $S^1$). 
As it stands, \eqref{d0emerge} misses the potential contribution from bound states of $D2$- and $NS5$-branes or equivalently of transversal (to $S^1$) $M2$- and $M5$-branes, respectively. 
In the following section, we will extend the GGV formula by including also these objects. 
We will see that this is mandatory in order to obtain all known instanton corrections to the $R^4$-term in lower space-time dimensions. We will interpret this result as evidence that the coupling \eqref{R4termd} can be obtained from an emergence computation if one includes all of the light states identified in \cite{Blumenhagen:2023tev,Blumenhagen:2023xmk}. Additionally, as we have already discussed in section \ref{regularizations}, while the authors of \cite{Green:1997as} chose to first Poisson resum over all integers in \eqref{d0emerge} and then subtract the $(0,\ldots,0)$ term, we will directly exclude this term from \eqref{d0emerge}.

Let us anticipate a key point that will be further discussed in the following sections. 
We are interested in integrating out  bound states of $D0$-, $D2$-, $NS5$-branes, or equivalently transverse $M2$-, $M5$-branes with Kaluza-Klein momentum along the eleventh direction. With the exception of pure $D0$-branes (Kaluza-Klein) towers, these states are not particle-like. As such, we lack the technology to integrate them out at one-loop in general. To circumvent this issue, we propose the following strategy. We compactify the critical type IIA theory, or equivalently the eleven-dimensional M-theory, down to the maximum number of spacetime dimensions allowing us to work with point-like particle states only. For example, we propose that we can capture the contribution of a $D0$/$D2$-brane tower by calculating a Schwinger-like integral in eight dimensions with only particle-like states running in the loop. These are arising from Kaluza-Klein states and fully wrapped $M2$-branes. A similar strategy applies to $M5$-brane contributions. Eventually, the different origin of the various terms in ten or eleven dimensions can be understood upon decompactification.

\subsection{\texorpdfstring{$R^4$}{TEXT}-term in ten dimensions}

Let us consider a compactification of M-theory on $S^1$, which corresponds to ten-dimensional type IIA at strong coupling.
It is known that the coefficient of the $R^4$-term contains only a
string tree-level and a one-loop contribution \cite{Green:1997as} that  we have already encountered,
\begin{equation}
\label{r410dknown}
a_{10}\simeq \frac{2 \zeta(3)}{g_s^2} +\frac{2\pi^2}{3}\,.
\end{equation}
Our purpose is to arrive at the same result from an emergence computation. 

For  $d=10$, the Schwinger integral \eqref{d0emerge} with only Kaluza-Klein states ($D0$-branes) is given by
\begin{equation}
\label{r410dschwinger}
a_{10}\simeq \frac{2\pi}{ r_{11}}  \sum_{m\neq 0} \int_0^\infty  \frac{dt}{t^{2}}  \;  e^{-\pi t \frac{m^2}{r_{11}^2}}\,,
\end{equation}
where we denote dimensionless radii as $r_i=R_iM_*$.
This integral is divergent for $t\to 0$ and we need to impose some regularization.
Following the same technique of \cite{Blumenhagen:2023tev}, which we have already successfully applied in the string theory limit, we introduce a regulator $\epsilon>0$, apply \eqref{int exp/t^2} and minimally subtract the divergent first term.
In this way we arrive at the only non-trivial term 
\begin{equation}
a_{10}\simeq \frac{2\pi^2}{r_{11}^3} \sum_{m\neq 0} m^2 \log\left(\frac{m^2}{\mu^2}\right),
\end{equation}
with $\mu^2=(\pi\epsilon)^{-1} r_{11}^2 \exp(1-\gamma_E)$. 
Using $\zeta(-2)=0$, the $\mu$-dependent term drops out and the rest can be calculated using \eqref{zeta id1}
\begin{equation}
\label{result10D}
a_{10}\simeq \frac{2 \zeta(3)}{r_{11}^3} =  \frac{2\zeta(3)}{g_s^2}\,.
\end{equation}
The Schwinger integral for the tower of $D0$-branes reproduces the tree-level term in the expansion of  \eqref{r410dknown} at small $g_s\,$.
This result was already obtained by GGV \cite{Green:1997as}, but it can now be interpreted as evidence for the emergent nature of M-theory interactions. 
However, the string one-loop term in \eqref{r410dknown} is missing in our computation above. 
In this respect, it is important to recall that GGV added an (infinite) constant $C$ that was fixed in a subsequent step to the correct value $2\pi^2/3$ by invoking T-duality in nine dimensions.
Therefore, the actual origin of that contribution in ten dimensions remains obscure and indeed it is a challenge of the microscopic theory to determine it. As we will see in section  \ref{sec:8d}, the resolution is similar to the already encountered  missing  $2\pi^2/3$ term in the real Schwinger integral for the perturbative string in ten dimensions.

\subsection{\texorpdfstring{$R^4$}{TEXT}-term in nine dimensions}

Let us consider M-theory on $T^2=S^1\times S^1$ which corresponds to the type IIA superstring in nine dimensions. Here the complete coefficient of the $R^4$-term in perturbative type IIA theory \cite{Green:1997di}, 
\begin{equation}
\label{r49dknown}
a_{9}\simeq \frac{2 \zeta(3)}{g_s^2} +\frac{2\pi^2}{3}\left(1+\frac{1}{\rho^2_{1}}\right)+ \frac{8\pi}{\rho_1 g_s} \sum_{m\ne 0}\sum_{n>0} \left\vert\frac{m}{n}\right \vert K_1\Big(2\pi|m| n\, \frac{\rho_{1}}{g_s}\Big)\, ,
\end{equation}
does receive corrections beyond one-loop, namely from Euclidean $D0$-instantons, denoted as $E\!D0$ in the following.  

For $d=9$, the Schwinger integral \eqref{d0emerge} with only $D0$-brane contributions is given by
\begin{equation}
    \label{r49dschwinger}
    a_{9}\simeq \frac{2\pi}{r_{11} r_{1}}  \sum_{(m,m_1)\neq(0,0)} \int_0^\infty  \frac{dt}{t^{3/2}}  \;  e^{-\pi t\left(  \frac{m^2}{r_{11}^2}+\frac{m_1^2}{r_{1}^2}\right)}\, .
\end{equation}
Once again we split the sum according to $\sum_{m,m_1\neq (0,0)}
=\sum_{m=0,m_1\neq 0}+\sum_{m\ne 0,m_1\in \mathbb Z}\,$ and we subtracted the term corresponding to $(0,0)$.
The first piece ($m=0$) can be treated in the same way as discussed in section \ref{regularizations}, using \eqref{int exp/t^(3/2)} to obtain
\begin{equation}
\label{r49dfirst}
 a^{(m=0)}_{9}\simeq \frac{2\pi^2}{3}\frac{1}{r_{1}^2 \, r_{11}} \,.
\end{equation}  
As for the second piece ($m\neq 0$), one first performs a Poisson resummation of the sum over $m_1\in\mathbb Z$ to obtain 
\begin{equation}
a^{(m\neq 0)}_{9}\simeq \frac{2\pi}{r_{11}} \sum_{m\ne 0} \sum_{m_1\in\mathbb Z}  \int_0^\infty  \frac{dt}{t^{2}} \;e^{ -\pi t \frac{m^2}{ r_{11}^2} -\frac{\pi}{t} m_1^2 r_{1}^2} .
\end{equation}
The $m_1=0$ term gives precisely the sum \eqref{r410dschwinger} from the previous case, that is equal to $2\zeta(3)/r_{11}^3$, while the remaining sum over $m_1\ne 0$ can be carried out using \eqref{besselrel} and results in
\begin{equation}
\label{r49dsecond}
a^{(m \neq 0)}_{9}\simeq \frac{2\zeta(3)}{r_{11}^3}+\frac{8\pi}{r^2_{11} r_{1}} \sum_{m\ne 0} \sum_{m_1>0} \left\vert{\frac{m}{m_1}}\right\vert K_1\left(2\pi  |m| m_1\, {\frac{r_{1}}{r_{11}}}\right)\,.
\end{equation}
Expressing the two terms \eqref{r49dfirst} and \eqref{r49dsecond}  in string units, one recovers almost all contributions in \eqref{r49dknown} except again for the one-loop term $2\pi^2/3$, which in the GGV approach appeared only after replacing an infinite constant by enforcing T-duality.
Note that in the decompactification limit, $r_{1}\to\infty$, one consistently recovers the ten-dimensional result \eqref{result10D}. 

The second term in \eqref{r49dsecond} has the correct dependence on $g_s$ to be interpreted as the contribution from $E\!D0$-brane instantons. 
It is a non-trivial fact that by integrating out states which are
light in the strong (string) coupling regime we recovered a result compatible
with string perturbation theory, i.e.~in the opposite regime 
$r_{11}\ll 1$. One could formally repeat the calculation by splitting the summation as $\sum_{m,m_1\in\mathbb{Z}}=\sum_{m_1=0,m\in\mathbb{Z}}+\sum_{m_1\neq0,m\in\mathbb{Z}}\,$, and then Poisson resumming over $m$. The result $\tilde{a}_9$ would match with \eqref{r49dfirst} and \eqref{r49dsecond} upon exchanging $r_{1}$ and $r_{11}$, namely 
\begin{equation}
 \tilde{a}_{9}\simeq\frac{2\pi^2}{3}\frac{1}{r_{11}^2r_{1}}+\frac{2\zeta(3)}{r_{1}^3}+\frac{8\pi}{r_{1}^2r_{11}}\sum_{m_1\neq0}\sum_{m>0}\left|\frac{m_1}{m}\right|K_1\left(2\pi m|m_1|\frac{r_{11}}{r_{1}}\right)   \,. 
\end{equation}
This would be an expansion valid in the $r_{11}\gg 1$ regime.
Given that \eqref{r49dschwinger} is symmetric under this exchange, we expect the action of a highly non-trivial map to be needed in order to recover \eqref{r49dknown} from the above formula.

Our calculation suggests a direct correspondence between the particle-like states we integrate out and the instanton corrections to the effective $R^4$-coupling.
Namely, the $D0$-branes with Kaluza-Klein momentum along the compact $I=1$ direction lead to $E\!D0$-brane instantons, whose one-dimensional world-volume is wrapped around the same (first) compact direction.
We denote this correspondence as
\begin{equation}
(D0;{\rm KK}_{(I)}) \longleftrightarrow   E\!D0_{(I)}\,.
\end{equation}
Note that the action of the corresponding instanton can be read-off from the argument of the modified Bessel function, as was already observed in \cite{Pioline:1997pu}.
On the left hand side we have the strongly coupled type IIA string theory, dual to M-theory on a circle, while on the right hand side we reorganize the couplings in a ``perturbative" expansion valid in the opposite regime $g_s\ll 1$.

\section{Emergence of \texorpdfstring{$R^4$}{TEXT}-terms in M-theory in \texorpdfstring{$d=8,7$}{TEXT}}
\label{sec:rtothe48d7d}

In this section, we evaluate the appropriate Schwinger integrals in eight and seven dimensions, where a couple of new interesting issues arise. 
In the strong coupling limit we are considering, we now need to include also contributions from wrapped $D2$-$D0$ bound states corresponding to transverse $M2$-branes carrying Kaluza-Klein momentum along the M-theory circle and giving rise to particle-like states in eight dimensions. 
How we proceed is inevitably very closely related to the previous work \cite{Kiritsis:1997em,Pioline:1997pu,Obers:1998fb,Obers:1999um}, where the guiding principle was U-duality of the final $R^4$-term so that also longitudinal $M2$- and $M5$-branes were integrated out. 
Nevertheless, we decided to present the computation in a pedagogical step by step manner, which makes evident where our novel regularization method of real Schwinger integrals enters and how the results are to be interpreted from the emergence point of view.

\subsection{\texorpdfstring{$R^4$}{TEXT}-term in eight dimensions}
\label{sec:8d}

Let us consider M-theory on $T^3$, which corresponds to the type IIA superstring in eight dimensions. Here, the coefficient of the $R^4$-term in perturbative type IIA theory is \cite{Green:1997as}
\begin{equation}
\label{r48dknown}
\begin{aligned}
a_8 &\simeq \frac{2\zeta(3)}{g_s^2}-\frac{2\pi}{\vartheta_{12}} \log\left(\rho_2^2 |\eta(iu)\eta(i\rho_1\rho_2)|^4\right) \\
&+ \frac{8\pi}{\rho_{1}g_s}\!\!\!\!\!\!\!\!\sum_{\substack{m>0\\ (m_1,m_2)\neq(0,0)}}\!\!\!\!\!\!\!\!\frac{m}{|m_1+im_2u|}K_1\left(2\pi\frac{\rho_{1}}{g_s}m|m_1+im_2u|\right)\,,
\end{aligned}
\end{equation}
where $u=\rho_2/\rho_1=r_2/r_1$.

For $d=8$, the Schwinger integral \eqref{d0emerge} with only Kaluza-Klein contributions is given by
\begin{equation}
\label{r48dschwinger}
a_{8,D0}\simeq \frac{2\pi}{r_{11} t_{12}}  \sum_{(m,m_1,m_2)\neq(0,0,0)} \int_0^\infty  \frac{d t}{t}\; e^{-\pi t \left(  \frac{m^2}{r_{11}^2}+\frac{m_1^2}{r_{1}^2}+\frac{m_2^2}{r_{2}^2}\right)}\,,
\end{equation}
where as usual the $(0,0,0)$ term has been subtracted and we denote $t_{ij}=r_ir_j$.
We split the sum as $\sum_{(m,m_1,m_2)\neq (0,0,0)} =  \sum_{m\ne 0,m_1,m_2\in \mathbb{Z}}+\sum_{m=0,(m_1,m_2)\neq (0,0)}$ and then apply the methods of the previous case.
After Poisson resummation over $m_1$ and $m_2$, the first piece ($m \neq 0$) is similar to the corresponding nine-dimensional integral and gives 
\begin{equation}
\label{r48dsecond}
a^{(m\neq 0)}_{8,D0}\simeq \frac{2\zeta(3)}{r_{11}^3}+\frac{8\pi}{r^2_{11}} \sum_{m>0}   \sum_{(m_1,m_2)\ne (0,0)} \frac{m}{ \sqrt{ m^2_1  r_1^2+m_2^2 r_{2}^2}}\; K_1\!\left(2\pi  \frac{m}{r_{11}} \sqrt{ m^2_1 r^2_1+m_2^2 r_{2}^2}\,\right)\,.
\end{equation}
Hence, we see that the Kaluza-Klein modes encode again the contribution of $E\!D0$ instantons wrapped on a $(m_1,m_2)$ 1-cycle of the rectangular $T^2$ transverse to the M-theory circle.
The second piece ($m=0$) is instead
\begin{equation}
\label{r48dfirst}
a^{(m=0)}_{8,D0}\simeq \frac{2\pi}{r_{11} t_{12}}  \sum_{(m_1,m_2)\neq(0,0)} \int_0^\infty  \frac{d t}{t}\; e^{-\pi t \left(\frac{m_1^2}{ r_{1}^2}+\frac{m_2^2}{ r_{2}^2}\right)}\,
\end{equation}
and it can be evaluated by the same procedure. We split the sum over $m_1$ into the  $m_1\neq 0$ and $m_1= 0$ terms. Starting from the first of these, after Poisson resummation over $m_2$ and applying the relation \eqref{besselrel}, we can write the result as
\begin{equation}
\label{8Dcomplexstruc}
\begin{aligned}
a^{(m=0\neq m_1)}_{8,D0}\simeq \frac{2\pi}{ r_{11}t_{12}} \left( \frac{\pi}{ 3} u + 4\!\! \sum_{m_1,m_2>0} \frac{1}{m_2} e^{-2\pi m_1 m_2 u} \right) = - \frac{2\pi}{r_{11} t_{12}} \log\left(|\eta(iu)|^4 \right)\,.
\end{aligned}
\end{equation}
It remains to compute the $m=m_1=0$ contribution, which is given by
\begin{equation}\label{log generating term}
a^{(m=0=m_1)}_{8,D0}\simeq\frac{2\pi}{r_{11}t_{12}}\sum_{m_2\neq 0}\int_0^\infty\frac{d t}{t}e^{-\pi t\frac{m_2^2}{r_2^2}}\,.   
\end{equation}
After using \eqref{int exp/t}, and choosing a regulator $4\pi e^{-\gamma_E}\epsilon/r_{11}$, we get 
\begin{equation}
\label{r48danomaly1}
a_{8,D0}^{(m=0=m_1)}\simeq-\frac{2\pi}{r_{11}t_{12}}\log (r_{11}r_2^2)\, . 
\end{equation}
This expression is key to eventually get a modular invariant result. However, let us remind that in GGV \cite{Green:1997as} such a term had somehow to be added by hand, while here we can reproduce it by integrating out a physical quantity and imposing the desired modular behavior. Without modular invariance there seems to be no obvious (to us) way to unambiguously determine the appropriate physical scale for the above integral, given the (non-)analytic properties of this $R^4$-term.
In fact, the role of this logarithmic term is very special to eight dimensions and it can also be related to the holomorphic anomaly. Of course, this is true also in the emergent string limit, but in that frame a modular invariant result incidentally arose more naturally.

However, this is not the end of the story, since in eight dimensions one also expects instanton contributions from fundamental string world-sheets wrapped on the $T^2$ orthogonal to the M-theory circle.   
Looking at the light towers of states with typical mass scale below or of the order of the species scale, 1/2-BPS $D2$-$D0$ bound states wrapping the $T^2$ are the obvious candidates for this purpose, being particle-like in eight dimensions. 
From the M-theory perspective, these are transverse $M2$-branes carrying Kaluza-Kein momentum only along the eleventh direction.

These 1/2-BPS bound states are subject to the general condition \cite{Obers:1998fb}
\begin{equation}
\sum_J n_{IJ} \, m_J=0\,,
\end{equation}
where $n_{IJ}$ is the wrapping number of the $M2$-brane along the 2-plane $(IJ)$ and $m_K$ the Kaluza-Klein momentum along the $K$-th internal direction (the notation is chosen to match conventions adopted later on). 
On $T^2$, this means that the transverse $M2$-brane cannot carry any longitudinal momentum. 
The masses of these particle-like states are
\begin{equation}
\mathcal{M}^2=n_{12}^2 t_{12}^2+ \frac{m^2}{r_{11}^2}\,,
\end{equation}         
where we remind that $t_{12} =r_1 r_2$ denotes the area of the $T^2$ in M-theory units, while $n_{12}$ is the wrapping number of a $D2$-brane. 
We extend the Schwinger integral to also include these $D2/D0$ bound states via
\begin{equation}
\label{r48dschwingerD2}
a_{8,D2/D0}\simeq \frac{2\pi}{r_{11} t_{12} }  \sum_{n_{12}\ne 0} \sum_{m\in\mathbb Z} \int_0^\infty  \frac{d t}{t} \; e^{ -\pi t \left(  n_{12}^2 t_{12}^2 +\frac{m^2}{r_{11}^2}\right)}\,.
\end{equation}
Up to a redefinition of the parameters, this is the same expression as $a^{(m=0\neq m_1)}_{8,D0}$ in \eqref{8Dcomplexstruc} and we thus find
\begin{equation}
\label{r48DD2final}
\begin{aligned}
a_{8,D2/D0}&\simeq \frac{2\pi}{r_{11} t_{12}} \bigg( \frac{\pi}{ 3} r_{11} \,t_{12} + 4\!\! \sum_{n_{12},m>0} \frac{1}{n_{12}} e^{-2\pi n_{12} m r_{11} t_{12}} \bigg) \\
&= - \frac{2\pi}{r_{11}t_{12}} \log\Big( |\eta(ir_{11} t_{12})|^4 \Big)\,.
\end{aligned}
\end{equation}
The exponential terms describe type IIA (fundamental) string instantons wrapped on the $T^2$, whereas the first term in \eqref{r48DD2final} is another perturbative contribution, which will be discussed in more detail in the following section. 
Putting our results together, namely \eqref{r48dsecond}, \eqref{8Dcomplexstruc}, \eqref{r48danomaly1} and \eqref{r48DD2final}, the complete eight-dimensional expression is
\begin{equation}
\begin{aligned}
a_8 &\simeq \frac{2\zeta(3)}{r_{11}^3} - \frac{2\pi}{r_{11}t_{12}} \log \Big(r_2^2\,r_{11} \big|\eta(iu)\,\eta(ir_{11}t_{12})\big|^4\Big)\\[0.1cm]
&+\frac{8\pi }{r_{11}^2 r_1}\!\!\!\!\!\!\!\!\sum_{\substack{m>0\\ (m_1,m_2)\neq(0,0)}}\!\!\!\!\!\!\!\!\frac{m}{|m_1+i m_2u|}K_1\left(2\pi\frac{r_{1}}{r_{11}}m\big|m_1+i m_2u\big|\right)\,,
\end{aligned}
\end{equation}
and, once expressed in type IIA variables it coincides with
\eqref{r48dknown}.

By including both $D0$-branes and  (transversely) wrapped $D2$-branes in the Schwinger integral, we recovered the only two kinds of instanton corrections compatible with the dimensionality of the present setup, namely $E\!D0$ and $EF\!1$ instantons. As showed explicitly with our calculation, they correspond to the following contributions in Schwinger-like integrals
\begin{equation}
\begin{aligned}
(D0;{\rm KK}_{(I)},{\rm KK}_{(J)}) &\longleftrightarrow E\!D0_{(I)}-E\!D0_{(J)}\,,\\ 
(D2,D0) &\longleftrightarrow  E\!F1\,.
\end{aligned}
\end{equation}

\subsection{Decompactification limit and stringy one-loop contribution}
\label{sec:decompact}

As we are going to argue, the eight-dimensional expression that we have just obtained is rich enough to contain the one-loop term, $2\pi^2/3$, that we could not determine in nine and ten dimensions in this M-theoretic limit. 
To understand this point, let us take a closer look at the contributions that are perturbative from the weakly coupled, large volume  type IIA perspective, i.e.~we suppress world-sheet and space-time instanton corrections.
These can be written as
\begin{equation}
a_8^{\rm pert.}\simeq  \frac{2\zeta(3)}{r_{11}^3}+ \frac{2\pi^2}{ 3} + \frac{2\pi^2}{3} \frac{1}{r_{11} r_{1}^2} +\frac{8\pi}{r_{11} r_1 r_2}   \sum_{m_1,m_2>0} \frac{1}{m_2} e^{-2\pi  m_1 m_2 u}\, ,
\end{equation}
where importantly the second term comes from  the non-exponential term
in \eqref{r48DD2final}, while the third one comes from the
non-exponential term in \eqref{8Dcomplexstruc}.
Let us decompactify the second direction, i.e.~take
$r_2\to\infty$. Then, the last term vanishes and we get
\begin{equation}
a_8^{\rm pert.}\simeq  \frac{2\zeta(3)}{r_{11}^3}+ \frac{2\pi^2}{3} + \frac{2\pi^2}{3} \frac{1}{r_{11} r_{1}^2}\simeq \frac{2 \zeta(3)}{g_s^2} +\frac{2\pi^2}{3}\left(1+\frac{1}{\rho^2_{1}}\right)\,.
\end{equation}
This is the full type IIA perturbative contribution to the coefficient of the $R^4$-term in nine dimensions, already encountered in \eqref{r49dknown}. In particular, it includes also the one-loop term, ${2\pi^2/3}$, which has been missing so far.
Further decompactifying the tenth direction yields the complete type
IIA result \eqref{r410dknown} in ten dimensions.

This is very  similar to the string theory calculation in section \ref{sec:R4stringth}. There, in ten dimensions our method was also missing the constant term, which however could be obtained in nine dimensions from integrating out string winding modes along the compact direction.
In string theory, implementing the 1/2-BPS condition via a lagrange multiplier and using modular invariance, one could define a complex version of the Schwinger integral that also gave the correct constant term, $2\pi^2/3$, in ten dimensions. The non-trivial question is whether one can generalize this to M-theory and thus provide a higher-dimensional definition of the Schwinger integrals that upon evaluation also yields the correct result in ten and nine dimensions.

In string theory the constant term $2\pi^2/3$ arose consistently in dimensions $d\le 9$ after summing over all string winding sectors. 
Hence, we conjecture a similar {\it self-consistency} of Schwinger integrals in M-theory:
\begin{quotation}
\noindent
Claim: {\it For toroidal compactifications of M-theory down to any spacetime dimension $d\le 8$, integrating out all particle-like Kaluza-Klein and transverse $M2$- and $M5$-brane bound states yields the constant term $2\pi^2/3$ contributing to the coefficient of $R^4$ at one-loop in $g_s = (M_* R_{11})^\frac32$.}
\end{quotation}
Further confirmation for this claim will be given in the following when looking at $d\leq 7$.
Let us emphasize that in view of the Emergence Proposal in M-theory this is  a non-trivial statement.
While the string one-loop term is embedded into M-theory by considering the Schwinger integral over the longitudinal $M2$-branes, it is not obvious at all that we can obtain the same result also from the Schwinger integral over only the light transverse ones.

\subsection{\texorpdfstring{$R^4$}{TEXT}-term in seven dimensions}
\label{sec_7dM}

Let us consider M-theory on $T^4$, which corresponds to the type IIA superstring in seven dimensions. Here, the $M5$-brane does not yet contribute to Schwinger integrals when following our strategy, as it does not give a particle-like state upon wrapping it on the $T^3$ transverse to the M-theory circle. Therefore, we have to find the 1/2-BPS bound states made from transverse $M2$-branes and Kaluza-Klein momentum, i.e.~from particle-like $D0/D2$-branes with Kaluza-Klein momentum along three toroidal directions transverse to $S^1$. 
Let us denote with $n_{IJ}$ the integer 2-form charge of a $D2$-brane wrapping a $T^2\subset T^3$ along the $(IJ)$-plane, and with $m_K$ the Kaluza-Klein momentum along the $K$-th direction.
Then, the 1/2-BPS conditions are \cite{Obers:1998fb,Obers:1999um} 
\begin{align}
\label{BPSone}
n_{[IJ}\,  n_{KL]}&=0\,,\\
\label{BPStwo}
\sum_{J=1}^{3}n_{IJ}\, m_J &=0\,,
\end{align}
with $I,J,K,L=1,2,3$. Notice that the first of these conditions is trivially satisfied in our seven-dimensional setup, while the second will give non-trivial constraints on momenta. 
Once translated to M-theory language, the expression \eqref{adGGVstring} for the coefficient of the $R^4$-coupling becomes 
\begin{equation}
\label{r47dschwinger}
a_{7}\simeq \frac{2\pi}{r_{11} t_{123}} \sum_{(m,m_I,n_{IJ})\neq (0,\vec{0},\vec{0})} \int_0^\infty  \frac{d t}{t^{\frac12}}\;   \delta({\rm BPS})\, e^{-\pi t\left(  \frac{m^2}{ r_{11}^2}+\sum_{I} \frac{m_I^2}{r_{I}^2}+\sum_{I,J} {n_{IJ}^2 t_{IJ} ^2}\right)},
\end{equation}
where the sums are over all three Kaluza-Klein momenta along $T^3$ and all three different planes wrapped by $D2$-branes and denoted as $t_{IJ}=r_Ir_J$. 
Moreover, $\delta({\rm BPS})$ implements the 1/2-BPS constraints \eqref{BPSone},\eqref{BPStwo} on the parameters $m_I$ and $n_{IJ}$.  
Solving these constraints leads to four different kinds of 1/2-BPS states, but we remind that we will subtract the term corresponding to $m=m_I=n_{IJ}=0$. In seven dimensions, we can organize all quantum numbers as
\begin{equation}
(n_{IJ},n_{IK},n_{JK},m_I,m_J,m_K;m)\in\mathbb{Z}^7\,,\quad I\neq J\neq K\in\{1,2,3\}\,,
\end{equation}
where we are distinguishing the momentum $m$ along the eleventh direction, which is not constrained by the BPS conditions. The four types of solutions of the BPS conditions read:
\begin{enumerate}
\item All $n_{IJ}=0$. These states carry unrestricted Kaluza-Klein momentum but no winding and we parametrize them as
\begin{equation}
\label{classone}
(0,0,0, m_{I},m_{J},m_{K};m)\,,\qquad\quad  m,m_{I,J,K}\in  \mathbb Z\,, \qquad I \neq J \neq K.
\end{equation}

\item One $n_{IJ}\ne 0$. Due to the 1/2-BPS constraint \eqref{BPStwo}, these states carry an unrestricted Kaluza-Klein momentum $m_K$ orthogonal to $n_{IJ}$ and we parametrize them as
\begin{equation}
\label{classtwo}
(n_{IJ}, 0,0,0,0, m_{K};m)\,,\qquad\quad   m,m_{K}\in  \mathbb Z\,, \qquad I \neq J \neq K.
\end{equation}

\item Two $n_{IJ} \neq 0$. Hence, we introduce coprime integers $\tilde{n}_{IJ},\tilde{n}_{IK}\ne 0$ such that $n_{IJ} = n\cdot \tilde n_{IJ}$, $n_{IK} = n\cdot \tilde n_{IK}$, with $n\in \mathbb{Z}$. 
The 1/2-BPS constraint \eqref{BPStwo} allows these states to carry an orthogonal Kaluza-Klein momentum and we parametrize them as ($I \neq J \neq K$)
\begin{equation}
\label{classthree}
(n\cdot \tilde{n}_{IJ}, n\cdot \tilde{n}_{IK},0,0, -p\cdot \tilde{n}_{IK},p\cdot \tilde{n}_{IJ}; m)\,,\qquad\qquad\qquad   n>0 ,\, m,p\in  \mathbb Z\,.
\end{equation}

\item Three $n_{IJ} \neq 0$. Hence, we introduce coprime integers $\tilde n_{IJ}, \tilde n_{IK}, \tilde n_{JK} \neq 0$ such that $n_{IJ} = n \cdot \tilde n_{IJ}$, $n_{IK} = n \cdot \tilde n_{IK}$, $n_{JK} = n \cdot \tilde n_{JK}$ with $n \in \mathbb{Z}$. 

The 1/2-BPS constraint \eqref{BPStwo} allows these states to carry an orthogonal Kaluza-Klein momentum and we parametrize them as ($I \neq J\neq K$)
\begin{equation}
\label{classfour}
(n\cdot \tilde{n}_{IJ}, n\cdot \tilde{n}_{IK},n\cdot \tilde{n}_{JK}, p\cdot \tilde{n}_{JK},-p\cdot \tilde{n}_{IK},p\cdot \tilde{n}_{IJ};m)\,,\qquad\quad   n>0 ,\, m,p\in  \mathbb Z\,.
\end{equation}  
\end{enumerate}
Observe that each kind of solution can be related to any of the previous kinds by setting any of the coprime numbers to zero and using the convention that ${\rm gcd}(1,0)=1$. 

After having solved the 1/2-BPS constraint explicitly, we can evaluate the Schwinger integral \eqref{r47dschwinger} for these four cases. Much like for the calculation in the perturbative string limit in eight dimensions, we treat cases $2$ to $4$ simultaneously by introducing a vector $\alpha=(\alpha_{12},\alpha_{23},\alpha_{13})$ encoding which wrapping numbers are non-vanishing.
Then, the evaluation of the corresponding Schwinger integrals  can be performed simultaneously using  
\begin{equation}
\label{r47dd2branes}
a^\alpha_{7,D2/D0}\simeq \frac{2\pi}{r_{11}t_{123}} \, 2^{|\alpha|}
\sum_{\tilde{\mathbf{n}}_\alpha>0} \sum_{n>0}\,\sum_{m,p\in\mathbb
  Z}\int_0^\infty \frac{dt}{ t^{\frac12}}\; \, e^{-\pi t\left(
    L_\alpha^2 n^2+ \frac{m^2}{r_{11}^2}+  \frac{p^2 L_\alpha^2}{t_{123}^2}\right)}\,,
\end{equation}
with the total $T^3$ volume $t_{123}=r_1 r_2 r_{3}$ and  
\begin{equation}
L_\alpha=\sqrt{\tilde{n}_{12}^2 t^2_{12}+\tilde{n}_{13}^2 t^2_{13}+\tilde{n}_{23}^2 t^2_{23}}\,.
\end{equation}  
The symmetry factor $2^{|\alpha|}$ arises after restriction to positive winding numbers. Starting from \eqref{r47dd2branes} and performing a Poisson resummation with respect to the integers $(m,p)$, we obtain two contributions. The first, $(m,p)\neq (0,0)$, is a sum over instantons given by 
\begin{equation}
\label{r47dd2branes1}
a^{\alpha,(1)}_{7,D2/D0}\simeq (2\pi) \, 2^{|\alpha |} \sum_{\tilde{\mathbf{n}}_\alpha>0} \sum_{n>0}\sum_{(m,p)\ne (0,0)}
\frac{e^{-2\pi n \sqrt{p^2 t_{123}^2 + m^2 r_{11}^2 L_\alpha^2}}}{\sqrt{p^2 t_{123}^2 + m^2 r_{11}^2 L_\alpha^2 }}\,.
\end{equation}
More precisely, when viewing the above exponent as an instanton
action, we understand that this can be interpreted as the contribution
from $E\!D2/E\!F1$ (bound state) instantons.

Let us compare the pure world-sheet instanton corrections to the string theory results from appendix \ref{7d pert terms}.
These arise from \eqref{r47dd2branes1} by setting $p=0$.
Then, including the prefactor $4\pi$, the three contributions with $|\alpha|=1$ are exactly equal to \eqref{1instantonstot}, the three contributions with $|\alpha|=2$ are equal to \eqref{2instantonstot} and the single contribution with $|\alpha|=3$ is equal to \eqref{3instantonstot}. 
We emphasize that this is a highly non-trivial consistency check, as in M-theory we are not integrating out fundamental strings, i.e.~longitudinal $M2$-branes, but bound states involving transverse $M2$-branes.

The second contribution arises from $(m,p)=(0,0)$ and  after employing $\zeta$-function regularization for the infinite sum over $n$, it can be brought to the simple form
\begin{equation}
\label{r47dd2branes2}
a^{\alpha,(2)}_{7,D2/D0}\simeq \frac{2\pi^2}{3}  2^{|\alpha|-1} \sum_{\tilde{\mathbf{n}}_\alpha>0}  1\,.
\end{equation}
Next, we sum over all $\alpha$ and regularize the sums over coprime numbers employing \eqref{regcoprime2}. 
Finally, using also \eqref{sumbinomial} we find
\begin{equation}
\begin{aligned}
\label{r47dd2branes2final}
a^{(2)}_{7,D2/D0}&= \sum_{|\alpha|=1}^3\binom{3}{|\alpha|} a^{\alpha,(2)}_{7,D2/D0}\\
&\simeq \frac{2\pi^2}{3}  \left( 3\cdot 1 \cdot 1 + 3\cdot 2 \cdot \left(-\frac{1}{2}\right)  + 1\cdot 4  \cdot \left(\frac{1}{4}\right)\right)  = \frac{2\pi^2}{3}  \,.
\end{aligned}
\end{equation}
Hence, the total constant $D2/D0$-brane contribution is the same as in eight dimensions and survives the decompactification limit. This supports the claim in section \ref{sec:decompact} and in particular the fact that the constant term $2\pi^2/3$ arises from integrating out extended objects, which in this limit and dimensions are transverse $M2$-branes. More in general, it provides a non-trivial check of the self-consistency of our approach. Indeed, the existence of new types of 1/2-BPS bound states involving $D2/D0$-branes and $\zeta$-function regularization work together in an intricate manner to produce the result. 
While the claim is that this behavior generalizes to all dimensions $d\le 5$, in the next section we will explicitly prove this statement for the six-dimensional case.

It remains to evaluate the Schwinger integral for the first set of 1/2-BPS states running in the loop, namely those parametrized as in \eqref{classone}. This is given by
\begin{equation}
a_{7,\rm KK}=\frac{2\pi}{t_{123}}\sum_{(m,m_I)\neq(0,\vec{0})}\int_0^\infty\frac{dt}{t^{1/2}}e^{-\pi t\left(\frac{m^2}{r_{11}^2}+\sum_{i=1}^3\frac{m_i^2}{r_I^2}\right)}
\end{equation}
and it is similar to the eight-dimensional computation, so that we only provide the final result.
After splitting $\sum_{(m,m_I)\neq(0,\vec{0})}=\sum_{(m_1,m_2,m_3)\neq(0,0,0)}+\sum_{m\neq 0}\sum_{m_1,m_2,m_3\in\mathbb{Z}}$, from the second sum ($m\neq 0$) we obtain the usual $E\!D0$ instanton contributions
\begin{equation}
\label{r47dKKonly}
a^{(1)}_{7,\rm KK}\simeq \frac{2\zeta(3)}{r_{11}^3}+\frac{8\pi}{r^2_{11}} \sum_{m>0} \sum_{(m_1,m_2,m_3)\ne (0,0,0)} \frac{m}{l}\; K_1\!\left(2\pi  \frac{m}{r_{11}} l\right)\,
\end{equation}
with
\begin{equation}
l=\sqrt{m^2_1 r^2_1+m^2_2 r^2_2 + m_{3}^2 r_{3}^2}\,.
\end{equation}
The remaining terms ($m=0$) only depend on the complex structure moduli.
Here we assume $r_1 < r_2 < r_3$ and continue the process of splitting the summations and performing Poisson resummations when possible, which leads us to\footnote{The terms (in \eqref{r47dKKonly}) that arise from sums over only one non-zero integer are known to have poles that cancel against each other \cite{Obers:1999um,Terras:1985}. This is compatible with our method as we find the following two diverging contributions that require regularization
$$\frac{2\pi}{r_{11}t_{12}}\zeta(1)\simeq\frac{2\pi}{r_{11}t_{12}}\int_{\epsilon_1}^\infty\frac{dx}{e^x-1}\simeq\frac{2\pi}{r_{11}t_{12}}\left(-\log{\epsilon_1}+\mathcal{O}(\epsilon_1)\right)$$
and
$$\frac{2\pi}{r_{11}t_{12}}\sum_{m_2\neq0}\int_{\epsilon_2}^\infty\frac{dt}{t}e^{-\pi tm_2^2/r_2^2}\simeq\frac{2\pi}{r_{11}t_{12}}\left(\gamma_E-\log(4\pi)+\log\left(\frac{\epsilon_2}{r_2^2}\right)+\mathcal{O}(\epsilon_2)\right)\,.$$
These can be completely cancelled via an appropriate choice of regulators, e.g.~via choosing $\epsilon_2=r_2^2\epsilon_1\,4\pi e^{-\gamma_E}$, in order to exclude any terms that would break modular invariance. Once again, the necessity to impose a particular regulator scale stems from the fact that we are dealing with poles obstructing analytic continuation. The expression \eqref{r48danomaly1} can also be recovered by scaling the regulators appropriately.
}
\begin{equation}
\begin{aligned}
\label{r47dKKonly2}
a^{(2)}_{7,\rm KK}\simeq \frac{2\pi^2}{3}\frac{1} {r_{11}\, r_1^2}&+ \frac{2\pi}{r_{11} r_1^2} \sum_{m_1 \neq  0}\sum_{(m_2,m_3) \neq (0,0)} \frac{1}{U}e^{-2\pi |m_1|U} \\
&+\frac{4\pi}{r_{11} t_{12}}\sum_{m_2 \neq 0} \sum_{m_{3}\neq 0}  K_0\left(2\pi \,|m_2 \,m_3| \frac{r_3}{r_2}\right)
\end{aligned}
\end{equation}
with
\begin{equation}
U^2=m^2_2 \left(\frac{r_{2}}{r_{1}}\right)^2 +m^2_3 \left(\frac{r_{3}}{r_{1}}\right)^2\,.
\end{equation} 
Since the starting expression is symmetric under permutations of the radii $r_1$, $r_2$, $r_3$, a highly non-trivial map of Bessel functions would be needed to relate the above result, obtained for $r_1 < r_2 < r_3$, to that corresponding to any other radii hierarchy. 
Finally, upon decompactifying the third direction, $r_3 \to \infty$, one consistently gets the eight-dimensional result \eqref{8Dcomplexstruc}.

\section{Emergence of \texorpdfstring{$R^4$}{TEXT}-terms in M-theory in \texorpdfstring{$d<7$}{TEXT}}
\label{sec:rtothe4d<7}

The analysis performed so far can be generalized to lower non-compact dimensions, i.e.~to compactifications on tori $T^{k+1}$ with $k\ge 4$.
The corresponding computations become increasingly involved, since there will be more and more 1/2-BPS bound states made from towers of states with typical mass scale not larger than the species scale. 
Evaluating the resulting contributions to the $R^4$-terms in their full glory and for all dimensions is beyond the scope of this paper.
In this section, we will content ourselves with just performing a couple of further tests for the emergence of the $R^4$-interaction in M-theory.

\subsection{Recovering the constant one-loop term in six dimensions}

The next setup to be considered is M-theory on $T^5$, which corresponds to the type IIA superstring in six-dimensions. 
This is the first instance where the 1/2-BPS condition \eqref{BPSone} involving only the $M2$-brane charges is not satisfied trivially, namely
\begin{equation}
\label{BPS-6d}
n_{[IJ} n_{KL]}=n_{12} n_{34} - n_{13} n_{24} +  n_{14} n_{23}=0\,.
\end{equation}  
In the following, we do not evaluate the complete Schwinger integral but rather focus on the  constant term, $2\pi^2/3$, which has become one of the main tests of our proposal. 
To stress this point once more, since we are not integrating out longitudinal $M2$-branes, i.e.~type IIA fundamental strings (which in section \ref{sec:R4stringth} we proved to be generating the constant term in the perturbative string limit), it is a non-trivial check of the Emergence Proposal that we can get this term also from integrating out transverse $M2$-branes, which are $D2$-branes in type IIA language.

The computation is involved but analogous to the seven-dimensional case  so that we can be brief and present the final result.
In table \ref{table_dim6}, we list all 1/2-BPS configurations of $D2$-branes. For each of them, the 1/2-BPS conditions \eqref{BPStwo} for the Kaluza-Klein momenta $m_1,\ldots, m_4$  admit a two-dimensional space of unconstrained integers, with respect to which we perform Poisson resummation. 
After setting them to zero, performing the integral and summing over the greatest common divisor of the winding numbers (index $n$ in \eqref{r47dd2branes}), one arrives at an expression with the familiar factor $2\pi^2/3$ multiplied by sums over $|\alpha|$ relatively coprime weighted by the symmetry factor $2^{|\alpha|-1}$ (compare with \eqref{r47dd2branes2})
\begin{equation}
(|\alpha|)_1=2^{|\alpha|-1} \sum_{\tilde{\mathbf{n}}_\alpha>0} 1 = (-1)^{|\alpha|-1}\,.
\end{equation}
Here and in the following, the index $1$ serves as a reminder that $|\alpha|$ refers to the number of coprime integers, ${\rm gcd}(\tilde n_1,\dots, \tilde n_{|\alpha|})=1$.
Then, for each case, the result of this computation is presented in table  \ref{table_dim6}, where we indicate also the final sums of coprimes that occur. One realizes that all these contributions precisely add up to the desired one-loop term $2\pi^2/3$. Only the case of six $D2$-branes is more involved than the seven-dimensional case we have discussed in detail, so let us briefly comment on it.

According to the general solution \eqref{Dioph 3vargen solution1}-\eqref{Dioph 3vargen solution3}, the 1/2-BPS condition \eqref{BPS-6d} is solved by
\begin{equation}
(n_{34},n_{24},n_{23}) =\left(\frac{\tilde{n}_{13}}{g_2} N + X_0 M, \frac{\tilde{n}_{12}}{g_2} N + Y_0 M, g_2 M\right) \,, \quad N,M \in \mathbb{Z}\,,
\end{equation}
where ${\rm gcd}(\tilde{n}_{12},\tilde{n}_{13},\tilde{n}_{14})=1$, $g_2={\rm gcd}(\tilde{n}_{12},\tilde{n}_{13})$ and $(X_0,Y_0)$ is a particular solution such that $\tilde{n}_{12}X_0-\tilde{n}_{13}Y_0=-\tilde{n}_{14}g_2$.
However, to have six $D2$-branes one has to ensure that all wrapping numbers are non-vanishing, so not all integers $N$ and $M$ are valid choices. Restricting to $M\neq 0$ ensures that $n_{23}\neq 0$ and gives rise to the contributions $(3)_1(3)_1$ for $N\neq 0$ and $(3)_1(2)_1$ for $N=0$ in table \ref{table_dim6}. We then subtract the contributions of those $(M,N)$ corresponding to $n_{34} = 0$ or $n_{24} = 0$, which are given by the solutions of two more Diophantine equations, leading to the last term in table \ref{table_dim6}.

\begin{table}[ht] 
\renewcommand{\arraystretch}{1.5} 
\begin{center} 
\begin{tabular}{|c|c|c|c|} 
\hline
\#$D2$ &  \# BPS conf. & coprime sums & $\times 2\pi^2/3$       \\
\hline \hline
$1$  &    $6$  &  $(1)_1=1$   &  $6$ \\
$2$  &   $12$  &  $(2)_1=-1$ & $-12$\\
$3$  &    $4$  &  {$(2)_1\cdot (2)_1=1$} & $4$\\
     &    $4$  &  $(3)_1=1$ & $4$\\
$4$  &    $3$  &  $(2)_1\cdot (2)_{1}=1$ & $3$ \\
$5$  &    $6$  &  $(3)_1\cdot (2)_{1}=-1$ & $-6$ \\
$6$  &    $1$  &  $(3)_1\cdot (3)_1+(3)_1\cdot (2)_1-2\times ( (3)_1\cdot (2)_1)=2$ & $2$ \\
\hline\hline
 total  & $36$  & - &  $1$ \\
\hline
\end{tabular}
\caption{Particle-like states running in the loop and contributing to constant terms in $d=6$ dimensions. The final scaling factor presented in the last column arises via multiplying the number of 1/2-BPS configurations for a given number of $D2$-branes with the corresponding numerical factor arising from the various sums.}
\label{table_dim6}
\end{center} 
\end{table}

\subsection{Contribution from  \texorpdfstring{$M5$}{TEXT}-branes in five dimensions}

For $d\le 5$ also wrapped transverse $M5$-branes, i.e.~type IIA $NS5$-branes,  give rise to particle-like contributions in Schwinger-like integrals. 
We do not intend to give the full result for these cases, but we just present a representative contribution in five dimensions that shows what kind of instanton corrections appear.
First, we recall the 1/2-BPS conditions for the bound states involving also transverse $M5$-branes. These are \cite{Obers:1998fb}
\begin{align}
\label{BPStM2}
&\sum_J n_{IJ} m_J=0\, \\
&n_{[IJ}\, n_{KL]} + \sum_P m_P \,n_{PIJKL}=0\,, \\
& n_{I[J} \,n_{KLMNP]}=0\,.
\end{align}
In five dimensions the third condition is always satisfied, while the second can be spelled out explicitly as
\begin{align}
\label{5dBPSm1}
n_{23} n_{45} -  n_{24} n_{35} +  n_{25} n_{34} + m_1 \nu_5  &=0 \,, \\
\label{5dBPSm2}
n_{13} n_{45} -  n_{14} n_{35} +  n_{15} n_{34} - m_2 \nu_5  &=0 \,, \\
\label{5dBPSm3}
n_{12} n_{45} -  n_{14} n_{25} +  n_{15} n_{24} + m_3 \nu_5  &=0 \,, \\
\label{5dBPSm4}
n_{12} n_{35} -  n_{13} n_{25} +  n_{15} n_{23} - m_4 \nu_5  &=0 \,, \\
\label{5dBPSm5}
n_{12} n_{34} -  n_{13} n_{24} +  n_{14} n_{23} + m_5 \nu_5  &=0 \,,
\end{align}
where we denoted the wrapping number of the transverse $M5$-brane as $\nu_5=n_{12345}$. 
These equations need to be solved together with the five 1/2-BPS conditions for Kaluza-Klein-momentum and transverse $M2$-branes, namely \eqref{BPStM2} with $I,J=1,\ldots,5$.
There exist a plethora of solutions to these ten relations, each giving a contribution to the overall Schwinger-like integral. 
Due to this complexity, we restrict ourselves to just a single typical contribution.

We turn on up to four $M2$-branes with wrapping numbers $n_{14}$, $n_{15}$, $n_{23}$, $n_{45}$, one $M5$-brane and three transverse Kaluza-Klein momenta, $m_1$, $m_4$, $m_5$ together with that along the eleventh direction, $m$.
We reparametrize the $M2$-brane wrapping number $n_{23}$ and the $M5$-brane wrapping number $\nu_5$ as $(n_{23},\nu_5)=N(\tilde{n}_{23},\tilde{\nu}_5)$, with ${\rm gcd}(\tilde{n}_{23},\tilde{\nu}_5)=1$ and $N>0$. We can allow for either of them to be zero by the convention that ${\rm gcd}(0,\pm 1)=1$. 
All other $M2$-brane wrapping numbers and the momenta $m_2,m_3$ are vanishing. Then, the 1/2-BPS conditions \eqref{5dBPSm2}, \eqref{5dBPSm3} are automatically solved, while \eqref{5dBPSm1}, \eqref{5dBPSm4} and \eqref{5dBPSm5} become linear Diophantine equations in two variables whose solution is given by
\begin{equation}
(n_{45},m_1)=P (-\tilde{\nu}_5,\tilde{n}_{23})\,,\quad  (n_{15},m_4)=Q (\tilde{\nu}_5,\tilde{n}_{23})\,,\quad  (n_{14},m_5)=R (-\tilde{\nu}_5,\tilde{n}_{23})\,,
\end{equation}
where $P,Q,R\in\mathbb Z$. 
Notice that any of these integers being vanishing gives still a solution. 
After a few simple manipulations, the contribution of this sector can be
expressed as the Schwinger-like integral
\begin{equation}
a_5^{\rm typ}\simeq\frac{2\pi}{r_{11} t_{12345}} \sum_{\tilde{n}_{23},\tilde{\nu}_5\in \mathbb{Z}}  \sum_{N>0} \sum_{P,Q,R,m\in\mathbb Z} \int_0^\infty dt\, t^{\frac12}\,  e^{-\pi t \left(N^2 t_{23}^2 L^2 + \frac{m^2}{ r_{11}^2} + \left( \frac{P^2}{ r_1^2} +\frac{Q^2}{ r_4^2} +\frac{R^2}{ r_5^2} \right)L^2\right)}
\end{equation}
with $L=\sqrt{\tilde{\nu}_5^2 t_{145}^2 +\tilde{n}_{23}^2}$ and $t_{12345}=r_1r_2r_3r_4r_5$. 
After performing Poisson resummation with respect to the integers $m,P,Q,R$, we again arrive at an integral with measure $\int dt/t^{3/2}$. 
Then, using \eqref{int exp/t^(3/2)}, we obtain the instanton sum 
\begin{equation}
a_5^{\rm typ}\simeq2\pi \sum_{\tilde{n}_{23},\tilde{\nu}_5\in \mathbb{Z}} \sum_{N>0} \sum_{(P,Q,R,m)\ne (0,0,0,0)} \frac{1}{S\, L^2}\,  e^{-2 \pi  N S}\,,
\end{equation}
with
\begin{equation}
\label{instantactionm5}
S=\sqrt{  P^2 t_{123}^2 + Q^2 t_{234}^2 +R^2 t_{235}^2 + m^2 \left(\tilde{n}_{23}^2 (r_{11} t_{23})^2+\tilde{\nu}_5^2 (r_{11} t_{12345})^2 \right) }\,.
\end{equation}
This may be interpreted as an Euclidean instanton action, namely that of a bound state of three $E\!D2$'s, one $E\!F1$ instanton and an instantonic object with action
\begin{equation}
S=  r_{11} t_{12345}  = r_{11} \left(\frac{M_*}{M_s}\right)^5\,  \vartheta_{12345}  =\frac{1}{g_s} \vartheta_{12345}\,,
\end{equation}
where $\vartheta_{12345}$ is the volume of the $T^5$ in string units. This is the action of an instantonic wrapped $E\!D4$-brane. Note that from the M-theory perspective, this is a longitudinal Euclidean $E\!M5$-brane, even though we integrated out particle-like transverse $M5$-branes.

In $d\le 4$, one can also combine the $M5$-brane with a transversal, internal  Kaluza-Klein-momentum $m_I$. 
Then the argument of the modified Bessel function becomes
\begin{equation}
S=  r_{I} \,t_{12345} =  \left(\frac{M_*}{M_s}\right)^6\,  \vartheta_{12345I}  =\frac{1}{ g^2_s}  \vartheta_{12345I}\,,
\end{equation}
which is indeed the action of a Euclidean $NS5$-brane. For completeness, we also observe that the three $E\!D2$-brane contributions in \eqref{instantactionm5} arise via the  combination of a particle-like $M5$-brane and a relatively longitudinal $M2$-brane.
Indeed, performing a Poisson resummation with respect to the $M2$-brane number
has led to an instanton action
\begin{equation}
S =  \frac{t_{IJKLM}}{t_{IJ}}  = t_{KLM}= \frac{1}{g_s}  \vartheta_{KLM}\,.
\end{equation}

Like in all previous examples, the zero mode $(P,Q,R,m)=(0,0,0,0)$ gives a contribution that can finally be expressed as
\begin{equation}
\label{m5zero}
a_5^{{\rm typ}(0)}= \frac{\pi^2}{3} \sum_{\tilde{n}_{23},\tilde{\nu}_5\in \mathbb{Z}} \frac{1}{\tilde{\nu}_5^2 t_{145}^2 +\tilde{n}_{23}^2 } =\frac{2}{ t_{145}} E_1(i t_{145})\,,
\end{equation}
where we used the definition of the $SL(2,\mathbb{Z})$ Eisenstein series 
\begin{equation}
E_s(\tau)= \sum_{(m,n)\neq (0,0)} \frac{y^s}{|m\tau+n|^{2s}}=\zeta(2s)\sum_{\substack{m,n\in \mathbb{Z}\\\ \rm{gcd}(m,n)=1}}\frac{y^s}{|m\tau+n|^{2s}}\,,
\end{equation}
with $y={\rm Im}(\tau)$ and $\rm{gcd}(0,\pm 1)=1$ is implied.
For $s=1$, this Eisenstein series has a simple pole, which can be
minimally subtracted using the (first) Kronecker limit formula
\begin{equation}
E_s(\tau)=\frac{\pi}{s-1} +2\pi\left(\gamma_E-\log 2 -{\frac12}   \log\left( y\, |\eta(\tau)|^4 \right)\right) + \mathcal{O}(s-1)\,,
\end{equation}
to obtain
\begin{equation}
a_5^{{\rm typ}(0)}= -\frac{2\pi}{t_{145}} \log\left( t_{145} |\eta(i  t_{145})|^4 \right)=\frac{2\pi^2}{3}  +\frac{8\pi}{t_{145}} \sum_{n_1,n_2>0} \frac{1}{n_2} e^{-2\pi n_1 n_2 t_{145}}+\ldots\,. 
\end{equation}
This reveals that $a_5^{(0)}$ contains a full sum of $E\!D2$ instantons and also a constant term. As a consequence, both transverse $M2$- and $M5$-branes sectors have a chance to contribute to the constant term. However, without performing the full calculation for any possible solution to the 1/2-BPS conditions, one cannot draw a conclusion regarding the exact origin of the constant term and, more specifically, the contribution of the $M5$-branes to it.

\section{Comments on the general structure of  \texorpdfstring{$R^4$}{TEXT}-terms in M-theory}
\label{sec:exceptional}

In this section we take a more general perspective and relate our work to the papers \cite{Kiritsis:1997em,Pioline:1997pu,Obers:1998fb,Obers:1999um,Bossard:2015foa,Bossard:2016hgy}.
Formally, the two approaches are very closely related, up to certain differences, which we will point out and interpret in the context of the Emergence Proposal.
Finally, we also make some comments about integrating out extended objects in M-theory.

\subsection{The exceptional  \texorpdfstring{$E_{k(k)}(\mathbb{Z}$)}{TEXT}  structure}
  
Similarly to us, the work \cite{Kiritsis:1997em,Pioline:1997pu,Obers:1998fb,Obers:1999um,Bossard:2015foa,Bossard:2016hgy} was motivated by the generalization of the GGV computation to include also higher dimensional branes of M-theory. 
The main difference is that they were imposing  the full U-duality group of toroidal compactifications of eleven-dimensional supergravity, so that they were integrating out in Schwinger-like integrals all wrapped brane states, i.e.~including also the longitudinal $M2$- and $M5$-branes.
In contrast, here we are interested in the decompactification limit of type IIA superstring theory, which means that we distinguished a very large
eleventh direction $r_{11}\gg 1$. Just like in string perturbation theory, we only include in Schwinger-like integrals those towers of light states with a typical mass scale not larger than the species scale. 
Nevertheless, a structure very similar to \cite{Kiritsis:1997em,Pioline:1997pu,Obers:1998fb,Obers:1999um,Pioline:2010kb,Bossard:2015foa,Bossard:2016hgy} appears in our approach as well. Eventually, we expect that both strategies agree and give the full expansion of the coefficients of the $R^4$-term, ours in the perturbative $r_{11}\gg 1$ limit, while theirs in the bulk (desert) of the moduli space $r_{11}\sim \mathcal{O}(1)$.

Let us collect all light transverse particle states that we integrate out in Schwinger-like integrals.
These form bound states together with the unrestricted $D0$-branes, i.e.~the Kaluza-Klein modes along the eleventh direction.
As in \cite{Obers:1999um}, they fit into representations of $E_{k(k)}(\mathbb{Z})$, though with the label $k=10-d$ reduced by one from the choice $k=11-d$ valid when including all (i.e.~also longitudinal) particle states.\footnote{To simplify notation, we will be denoting $E_{k(k)}(\mathbb{Z})$ as $E_{k(k)}$.}
As usual, we define $E_{2 (2)}=SL(2)$, $E_{3 (3)}=SL(3)\times SL(2)$, $E_{4 (4)}=SL(5)$ and $E_{5 (5)}=SO(5,5)$.
To be concrete, let us collectively denoted the transverse Kaluza-Klein momenta $m_I$, $I\in\{1,\ldots, k\}$ and the various $M$-brane wrapping numbers as 
\begin{equation}
(N^A)=\left(m_I, n^{IJ}, n^{IJKLM}\right)\,,
\end{equation}     
so that the mass matrix for 1/2-BPS states is given by
\begin{equation}
{\cal M}={\rm diag}\bigg\{\frac{1}{r_I^2}, t_{IJ}^2, t_{IJKLM}^2\bigg\}\, .
\end{equation}
As throughout the rest of the paper, we made the simplifying assumption of a rectangular torus with vanishing axionic fields.  Turning them on leads to off-diagonal components of the mass matrix ${\cal M}$, which will provide the expected ``complexification'' of the perturbative and instanton expansion of the coefficient of the $R^4$-term.
The 1/2-BPS conditions involving Kaluza-Klein modes and the transverse $M2$- and $M5$-brane wrapping numbers read \cite{Obers:1998fb}
\begin{align}
&n^{IJ} m_J=0\,,\hspace{3.25cm} \#=k,\\
&n^{[IJ}\, n^{KL]} + m_P \,n^{PIJKL}=0\,, \hspace{0.4cm} \#={\binom{k}{4}},\\
& n^{I[J} \,n^{KLMNP]}=0\,, \hspace{2cm}  \#=k \times \binom{k}{6},
\end{align}
where we also indicate the number of different equations.
The first condition says that the momentum has to be orthogonal to the world-volume of the $M2$-brane. For vanishing $M5$-brane wrapping number, the second conditions means that the matrix $n^{IJ}$ has rank two.

In table \ref{table_particlesinE}, for $k\leq 6$ we list all particle states and how they fit into representations $\Lambda_{E_k}$ of $E_{k(k)}$, as well as the representations $\lambda_{E_k}$ of the 1/2-BPS conditions (the latter turn out to coincide with the representation of the string multiplet \cite{Pioline:2010kb}).\footnote{Note that for $k>6$ also transversely wrapped Kaluza-Klein-monopoles of M-theory would enter.}
\vspace{0.3cm}
\begin{table}[ht] 
\renewcommand{\arraystretch}{1.5} 
\begin{center} 
\begin{tabular}{|c|c|c|c|c|c|} 
\hline
$d$ & $k$ & Particles $SL(k)$ reps. &  $E_{k(k)}$ &  $\Lambda_{E_k}$  &  $\lambda _{E_k}$ (1/2-BPS)     \\ 
\hline 
\hline
9 & 1 & $[1]_p$ & 1 & 1 &0\\
8 & 2 & $[2]_p + [1]_{M2}$  &  $SL(2)$ & 3 & 2  \\
7 & 3 & $[3]_p + [3]_{M2}$  &  $SL(3)\times SL(2)$ & (3,2) & (3,1)  \\
6 & 4 & $[4]_p + [6]_{M2}$  &  $SL(5)$ & 10 & 5  \\
5 & 5 & $[5]_p + [10]_{M2}+[1]_{M5}$  & $SO(5,5)$  &  16 & 10  \\
4 & 6 & $[6]_p + [15]_{M2}+[6]_{M5}$  &   $E_6$ & 27 & 27  \\
\hline      
\end{tabular}
\caption{Particle states, 1/2-BPS conditions and their $E_k$ representations. Here, $[k]_p$ denotes the Kaluza-Klein-momenta along the $k$ transverse directions.}
\label{table_particlesinE}
\end{center} 
\end{table}
\noindent
By adopting a notation similar to that of (constrained) generalized Eisenstein series in \cite{Obers:1999um}, we can define the full $R^4$-term $a_d$ in $d$ dimension as
\begin{equation}
\label{fullr4termdef}
{\cal E}^{E_{k(k)}}_{\Lambda_{E_k}\oplus 1, s={\frac k2}-1}=\frac{2\pi}{r_{11}{\cal V}_{k}}   \hat{\sum_{N^A,m\in\mathbb Z}} \int_0^\infty  \frac{dt}{t} \frac{1}{t^{{\frac k2}-1}}\; \delta(\lambda_{E_k}) \;  e^{-\frac{\pi}{t}\, \left( N^A  {\cal M}_{AB}  N^B+ \,\frac{m^2}{r_{11}^2}\right)}\,,
\end{equation}
where the contribution with  all integers being vanishing has been subtracted as usual (hat on the sum symbol) and by $\Lambda_{E_k}\oplus 1$ we are including the singlet associated to the eleventh Kaluza-Klein mode.

The form \eqref{fullr4termdef} is analogous to the definition used in the early work \cite{Obers:1999um} and shows that, up to some normalization factors, ${\cal E}^{E_{k(k)}}_{\Lambda_{E_k}\oplus 1,s=k/2-1}$ is equal to a constrained Eisenstein-like series of $E_{k(k)}$ in the (reducible) representation $\Lambda_{E_k}\oplus 1$  at order $s=k/2-1$. 
The main difference with respect to \cite{Obers:1999um} is that there all  wrapped $M$-branes, including also the longitudinal ones, were integrated out.
Hence,  the $R^4$-coefficient was conjectured to be expressed (in the particle multiplet) as
\begin{equation}
a_d={\cal E}^{E_{k+1(k+1)}}_{\Lambda_{E_{k+1}}, s={\frac k2}-1}\,,
\end{equation}
with still  $d+k=10$ for $k>2$.
This was reflecting the full U-duality group of a toroidal compactification of M-theory.
In our large $r_{11}$ limit, $T$-duality in the eleventh direction maps the setup to the perturbative type IIA theory with $r_{11}\ll 1$.
Hence, in the asymptotic decompactification limit the U-duality group $E_{k+1(k+1)}$ is broken to $E_{k(k)}$.

This makes it evident that physically the difference between our approach and \cite{Obers:1999um} is that we are working consistently in the perturbative
decompactification limit where we only integrate out the perturbative states, i.e. transverse $M$-branes, whereas in  \cite{Obers:1999um}  the Schwinger integral involves all states. 
From a swampland perspective, this means that their computation is relevant in the regime where  no hierarchy between light and heavy towers exists, i.e.~all states are parametrically of the same mass. 
This regime with $r_{11}=\mathcal{O}(1)$ has been called the desert in \cite{Long:2021jlv}. 
Assuming that also in that region the 1/2-BPS saturated $R^4$-coefficient can be computed via a Schwinger integral over all states, consistency would lead us to the intriguing relation
\begin{equation}
\label{eisenidentity}
{\cal E}^{E_{k(k)}}_{\Lambda_{E_k}\oplus 1,s= {\frac k2}-1}= {\cal E}^{E_{k+1(k+1)}}_{\Lambda_{E_{k+1}}, s={\frac k2}-1}\,,
\end{equation}
between two constrained Eisenstein series which are a priori different.
As we have seen, the Schwinger integral on the left-hand side is missing the constant $2\pi^2/3$ term  for $k\le 1$ so that  as a true identity, \eqref{eisenidentity} can only be correct for $k\ge 2$. 
As we have argued, we can get this term upon going to eight dimensions and then decompactifying so that we just define the left-hand side of \eqref{eisenidentity} to include this additional term for $k=0,1$.

\begin{figure}[ht]
\centering
\vspace{-2.3cm}
\includegraphics[width=\textwidth]{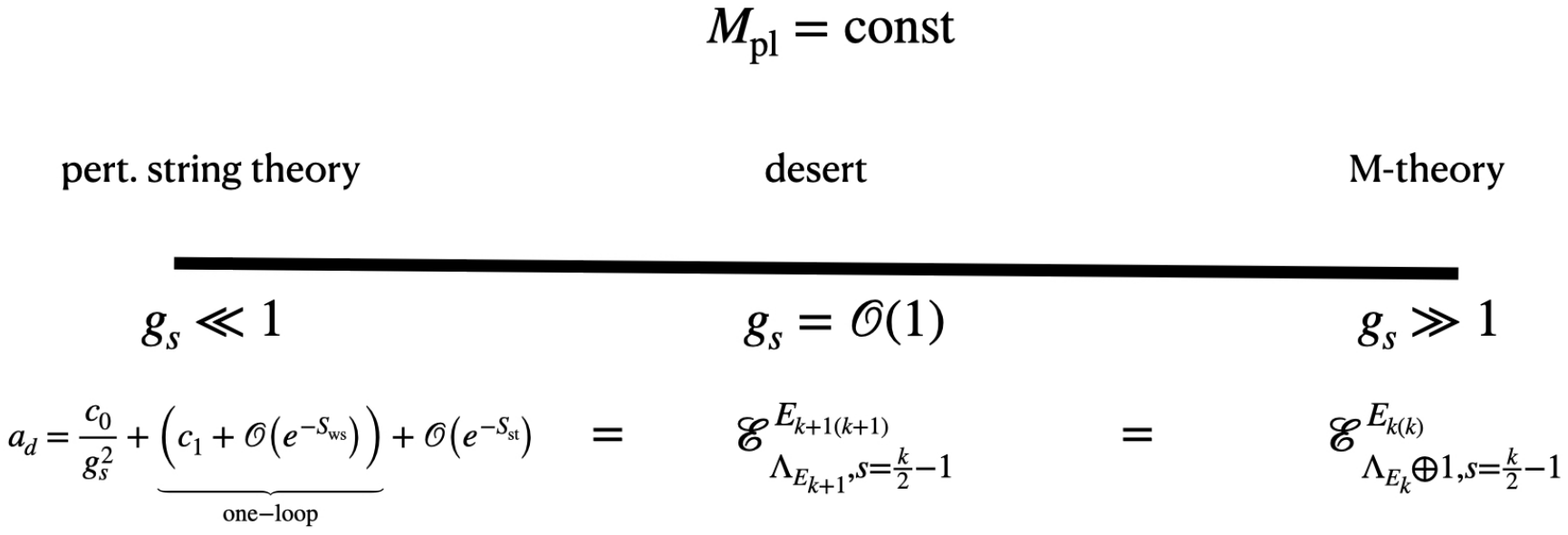}
\vspace{-3.5cm}
\caption{Schematic view of the dilaton moduli space.}
\label{fig_moduli}
\end{figure}

Hence, our swampland considerations based on infinite distance limits and the species scale have guided us to a testable mathematical relation. 
Using the here employed methods, we will provide a proof of this relation
for $d\ge 8$. A more general proof for $d\ge 4$ is already available
in \cite{Bossard:2015foa,Bossard:2016hgy}.\footnote{We are indebted to   Guillaume Bossard and Axel Kleinschmidt for pointing this out to us. Indeed, performing a Poisson resummation with respect to the Kaluza-Klein momentum $m$ in \eqref{fullr4termdef}, after a few manipulations one obtains the right hand side of the relation (3.18) in  \cite{Bossard:2016hgy} (for vanishing axions).}
In figure \ref{fig_moduli}, we show the physical picture behind the relation \eqref{eisenidentity}.

\subsection{Evidence for correspondence of Eisenstein series}

The equality \eqref{eisenidentity} is trivially satisfied in ten dimensions, as there only the Kaluza-Klein modes along the eleventh direction contribute on both sides.
Close inspection of the results already obtained allows us to prove \eqref{eisenidentity} in nine and eight dimensions. 
In nine dimensions, including all particle excitations amounts to adding the contribution of an $M2$-brane wrapping the directions $1$ and $11$ with no allowed momentum.
This contribution  to the right-hand side of \eqref{eisenidentity} was also obtained in
\cite{deWit:1999ir} and reads
\begin{equation}
a_{9}\simeq \frac{2\pi}{r_{11} r_{1}}  \sum_{n\neq 0}   \int_0^\infty  \frac{dt}{t^{3/2}}  \;  e^{-\pi t   n^2 (r_{11}  r_1)^2}    =\frac{2\pi^2}{3}\,.
\end{equation}
As expected, this is the string winding mode contribution \eqref{string9dcontribution}  written in  M-theory units.
Recalling that this term is included into the definition of the left-hand side of \eqref{eisenidentity}, that relation is proven in nine dimensions.

In eight dimensions, the pure Kaluza-Klein sector is the same on the left and the right hand side of \eqref{eisenidentity}.
Recall that on the left-hand side there is only a single transverse $M2$-brane contribution, namely \eqref{r48DD2final}. 
On the contrary, on the right hand side one has to sum over various $M2$-bound states, which are very similar to the transverse  $M2$-bound states for $d=7$.
In fact, formally we only have to replace  $r_3 \rightarrow r_{11}$  in the Schwinger-like integrals in section \ref{sec_7dM}, remove the  sum over the Kaluza-Klein-momentum $m$ in the eleventh direction and adjust the measure of integration appropriately.
In this manner, the result \eqref{r47dd2branes} can be recast as
\begin{equation}
\tilde{a}^\alpha_{8,D2} \simeq \frac{2\pi}{r_{11}t_{12}} \, 2^{|\alpha|} \sum_{\tilde{\mathbf{n}}_\alpha>0} \sum_{n>0}\,\sum_{p\in\mathbb Z}\int_0^\infty \frac{dt}{t} \; \, e^{-\pi t \left( L_\alpha^2 n^2 +\frac{p^2L_\alpha^2}{t_{12}^2 r_{11}^2}\right)}  \,,
\end{equation}
with $L_\alpha = \sqrt{\tilde{n}_{12}^2 t_{12}^2 + n_{1(11)}^2 t_{1(11)}^2 + n_{2(11)}^2 t_{2(11)}^2}$.
We can now proceed as usual and perform a Poisson resummation with respect to  $p$, carry out the integral with \eqref{besselrel} and arrive at
\begin{equation}
\tilde{a}^\alpha_{8,D2} \simeq  2^{|\alpha|-1}\left(\sum_{\tilde{\mathbf{n}}_\alpha>0} 1\right) \left( \frac{2\pi^2}{3}+ 8\pi \, \sum_{n,p>0}\, \frac{ e^{-2\pi n p\, r_{11} t_{12}}  }{  p \,r_{11} t_{12} } \right)\,.
\end{equation}
Since this only depends on $|\alpha|$, we can  employ  \eqref{regcoprime2} and  \eqref{sumbinomial} to sum over all $\alpha$ and get precisy the desired result
\eqref{r48DD2final}. This proves the identity \eqref{eisenidentity}
in eight dimensions.

Moreover, we  notice that the one-loop (in $g_s$) contribution is correctly reproduced. 
Since longitudinally wrapped $M2$-branes become fundamental strings of type IIA, the right hand side of \eqref{eisenidentity} contains these perturbative terms. As we have already discussed, for the left hand side this is not clear at all and their appearance implies a non-trivial check of the relation \eqref{eisenidentity}.
In the course of this paper, we have confirmed their appearance a couple of times for $d\ge 6$. 

Another non-trivial check is that our restricted sum, at least in principle,  induces all expected space-time instantons.
From the discussion in the previous sections, we found the  correspondence between particles in the loop and instantons as summarized in table \ref{table_corresp}.
Here  e.g.~$(D0,{\rm KK}_{(K)})$  means that we focus on this contribution to \eqref{fullr4termdef}, perform a Poisson resummation on the Kaluza-Klein mode in the $K$-direction and apply the formula \eqref{besselrel}.
The argument of the Bessel functions allows us to read-off the corresponding instanton action.
This shows in particular that even though we are not integrating out particle states resulting from wrapped  fundamental strings or $D4$-branes, i.e.~longitudinal $M2$- and $M5$-branes,  their instantons are nevertheless arising from  integrating out transverse $M2$- and $M5$-branes.

\vspace{0.3cm}
\begin{table}[ht] 
  \renewcommand{\arraystretch}{1.5} 
  \begin{center} 
    \begin{tabular}{|c|c|} 
      \hline
    Particle states &   Instantons        \\
      \hline \hline
      $(D0,{\rm KK}_{(K)})$   & $E\!D0_{(K)}$ \\
      $(D2_{(IJ)},{\rm KK}_{(K)})$   & $E\!D2_{(IJK)}$ \\
        $(N\!S5_{(IJKLM)},{\rm KK}_{(N)})$   & $E\!N\!S5_{(IJKLMN)}$ \\
      \hline
      $(D2_{(IJ)},D0)$   & $E\!F1_{(IJ)}$ \\
      $(N\!S5_{(IJKLM)},D0)$   & $E\!D4_{(IJKLM)}$ \\
      \hline
      $(N\!S5_{(IJKLM)},D2_{(LM)})$   & $E\!D2_{(IJK)}$ \\
      \hline
      \end{tabular}
          \caption{Particle (in loop) - instanton correspondence for
            elementary states.}
    \label{table_corresp}
  \end{center} 
\end{table}

\subsection{Schwinger integrals for extended objects}

Finally, let us speculate on an important issue on which the investigation of this paper might provide a new look. 
As manifest in our analysis, the evaluation of the naively UV-divergent Schwinger-like integrals is still at an initial stage, and we are dealing with the divergence at $t=0$ by imposing minimal subtraction and $\zeta$-function regularization.
As discussed in section \ref{sec2:preliminaryGGVst}, we could handle the UV divergences in string one-loop amplitudes in the same fashion, but the price to pay is that one is missing the constant contribution $2\pi^2/3$  in ten dimensions for string theory, and in ten and nine dimensions for M-theory.
These contributions are correctly obtained only by the finite and well-defined string theoretic one-loop amplitude \eqref{ad pert st general}, where one integrates the complex structure modulus $\tau=\tau_0+i t$ of the torus over the fundamental domain of the modular group $SL(2,\mathbb Z)$. Here, $\tau_0$ could be interpreted as a Lagrange multiplier for the single 1/2-BPS constraint or the level matching condition, respectively.

A natural question is whether one can proceed in a similar manner for the M-theory Schwinger integrals and define what one means by integrating out extended objects, such as $M2$- and $M5$-branes.
An essential difference with respect to the string theory case is that the number of 1/2-BPS conditions increases with the number $k$ of compactified directions. 
Concretely, the latter transform in the representation $\lambda_{E_k}$ so that, after imposing them via Lagrange multipliers, we arrive at an integral over ${\rm dim}(\lambda_{E_k})+1$  dimensions.
The question is then whether this can be interpreted as the moduli space ${\cal M}_k$ of an $M2$-$M5$ world-sheet of dimension 
\begin{equation}
{\rm dim}({\cal M}_k)= {\rm dim}(\lambda_{E_k})+1\,.
\end{equation}
    
At the moment, this is not at all obvious to us.
Apparently, for M-theory the story is more involved than for string theory and it could very well be that a first quantized world-sheet approach is not the right way to proceed. 
This point is also supported by the BFSS matrix model \cite{Banks:1996vh}, where the transverse $M2$- and the longitudinal $M5$-brane are emerging from the interactions of a matrix-valued quantum mechanics, which should rather be considered as a second quantized theory. 
Notice that this aforementioned increasing complexity  with the number of compact dimensions is very similar to what one encounters in generalizing Double Field Theory to Exceptional Field Theory and also in the BFSS matrix theory \cite{Banks:1996vh} approach to M-theory.
Hence, it appears to be an inherent property of M-theory, at least when truncated to the 1/2-BPS sector.

\section{Conclusions}
\label{sec:conclusion}

In the present work we provided evidence for the realization of an M-theoretic Emergence Proposal, which first appeared in the analysis of a certain  decompactification limit in the vector multiplet moduli space of type IIA compactified on a Calabi-Yau threefold to four dimensions \cite{Blumenhagen:2023tev}.
Employing in a broader context the same regularization procedure of \cite{Blumenhagen:2023tev} for real Schwinger-like integrals, we evaluated here  1/2-BPS saturated $R^4$-couplings for type IIA string theory in various dimensions. 
This was achieved by extending the original computation of Green-Gutperle-Vanhove \cite{Green:1997as} for toroidal compactifications of M-theory to include also light particle-like towers of transverse $M2$- and $M5$-branes  wrapped on appropriate cycles. 

We began our discussion with the emergent string limit, $g_s \ll 1$, to present the technical aspects of our regularization procedure in a familiar framework. 
By integrating out Kaluza-Klein and winding modes, the Schwinger integral gave the full finite one-loop result for $d\leq 9$ including all world-sheet instanton corrections. 
In this limit, one had to impose only a single BPS condition, relating the Schwinger integral to the UV-finite integral over the modular parameter of the string world-sheet. 
Importantly, the tree level terms could not be recovered in such a way, thus confirming the findings of \cite{Blumenhagen:2023tev} regarding the limitations of the Emergence Proposal in emergent string limits.

Next, we applied the Schwinger integral formalism to the M-theory
limit, building upon and extending the work of Green-Gutperle-Vanhove \cite{Green:1997as}. In our approach, transverse $M2$- and $M5$-branes, with Kaluza-Klein momentum along the eleventh and other compact directions, were the light degrees of freedom. In this setup, the appropriate Schwinger integral in principle gave the complete answer, namely tree-level, one-loop and instanton contributions from the weakly coupled point of view.
However, as for the Schwinger integral in the emergent string limit in ten dimensions, we missed the constant one-loop term in $d=10,9$, where only the Kaluza-Klein modes are particle-like. 
By further compactification to $d=8,7,6$ we were able to interpret the origin of this term as coming from the contributions of transverse $M2$-branes carrying Kaluza-Klein momentum, which survive the decompactifications.
Due to the increasing complexity of the BPS conditions in toroidal compactifications of M-theory down to lower dimensions, we did not yet manage to evaluate the complete Schwinger integrals in cases where also particle-like wrapped $M5$-branes contribute.
Since in the infinite distance limit, $r_{11}\gg 1$, the mass scale of the type IIA fundamental string, i.e.~the longitudinal $M2$-brane, is above the species scale $M_*$, it is a non-trivial test that one nevertheless recovers the string one-loop corrections. We provided explicit checks of this in $d\geq 7$.
The obtained results further support our claims for both the validity of our physics motivated regularization procedure and the notion of exact emergence in M-theory.

We pointed out that our approach is intricately related to the one of  Kiritsis-Obers-Pioline \cite{Kiritsis:1997em,Pioline:1997pu,Obers:1998fb,Obers:1999um}, who were investigating the behavior of the $R^4$-couplings based on the full U-duality group in each dimension. 
They realized that the appearing Schwinger integrals can be considered as constrained Eisenstein series of the related U-duality group.
We emphasized that there are two differences between our approach and theirs, one of rather technical and the other of conceptual nature.
First, they were employing analytic continuation for the simpler unconstrained Eisenstein series and were assuming it could possibly also be employed as a regularization method for the constrained divergent Eisenstein series, whereas we followed the more physical approach of minimal subtraction and $\zeta$-function regularization.
The conceptual difference between our work and theirs is
that we restricted the Schwinger integrals to  contributions from light towers of states in decompactification limits of type IIA string theory. 
From the point of view of the distance conjecture and the Emergence Proposal those are the towers whose typical mass scales are not parametrically larger than the species scale. 
Contrarily, in the work of Kiritsis-Obers-Pioline all towers of states where integrated out.
We explicitly showed that both approaches led to the same result
for $d\ge 8$,  giving rise to a  correspondence between the constrained Eisenstein series ${\cal E}^{E_{k(k)}}_{\Lambda_{E_k\oplus1},s=\frac{k}{2}-1}$ and  ${\cal E}^{E_{k+1(k+1)}}_{\Lambda_{E_{k+1}},s=\frac{k}{2}-1}$.
Intriguingly, this mathematical  relation was  already proven  for $d\ge 4$ in \cite{Bossard:2015foa,Bossard:2016hgy}. We find this to be a nice example of how swampland-related arguments, enriched by the notion of species scale, can suggest concrete and non-trivial mathematical statements.

The physical interpretation behind it  is the following: whereas our approach computes the coefficient of the $R^4$-term in the perturbative M-theory limit, their computation has its validity in the so-called desert region $g_s\sim\mathcal{O}(1)$, where all towers of states have the same parametric mass scale.
From this perspective, both approaches should in the end give the same result.

All the evidence that we have collected so far for the relevance of the Emergence Proposal for M-theory stems from treatable case of 1/2-BPS saturated terms in the effective action, like the $R^4$-terms here discussed or the topological amplitudes  in  compactifications of type IIA to four-dimensions with $\mathcal{N}=2$ supersymmetry \cite{Blumenhagen:2023tev}.
An obvious direction for future research would be to apply our techniques to other protected couplings. 
In this respect, one could look first at other higher derivative couplings, such as $D^4R^4$ ($1/4$-BPS) and $D^6R^4$ ($1/8$-BPS), before addressing the considerably more complicated problem of studying non-supersymmetric interactions. In relation to the species scale, these couplings have been recently investigated in \cite{Castellano:2023aum}. 
More in general, an important  question is whether emergence is just a special aspect of such a simplified set-up or whether it is a general property of M-theory. 
In the latter case, also the ten-dimensional tree-level kinetic terms have to arise from quantum effects, presumably together with  space-time itself.
However, the high degree of supersymmetry forbids such couplings to be generated at one-loop.
Does this mean that the Emergence Proposal is not realized beyond the 1/2-BPS level?
This conclusion might be premature, as for such a non-supersymmetric coupling  also the excitations of M-theory will enter. 
As we have learned from matrix theory, these are not simply the first quantized excitations of $M2$- and $M5$-branes, which would be the direct analogues to string excitations.
Hence, one needs to understand the generation of such couplings 
in M-theory at the quantum mechanical level.
Let us close with mentioning that indeed it was shown that the leading order (in a relative velocity expansion) classical interaction of two (super-)gravitons in supergravity is generated at the one-loop level in  the BFSS matrix model \cite{Douglas:1996yp, Banks:1996vh}. This fact is very much in agreement with the M-theoretic Emergence Proposal.

\paragraph{Acknowledgments.}
We thank Carlo Angelantonj, Guillaume Bossard, Elias Kiritsis, Axel Kleinschmidt, Dieter L\"ust, Eran Palti, Timo Weigand and Max Wiesner for useful discussions. The work of R.B.~and A.G.~is funded by the Deutsche Forschungsgemeinschaft (DFG, German Research Foundation) under Germany’s Excellence Strategy – EXC-2094 – 390783311.


\appendix

\section{Perturbative terms in seven dimensions}
\label{7d pert terms}
In this appendix, we compute the world-sheet instanton corrections to the coefficient $a_7$ of the $R^4$-term in seven dimensions. Since these are perturbative in $g_s$, we can employ the methods presented in section \ref{string limit}. 
For this purpose, we consider the contributions from one, two and three non-vanishing string winding numbers separately and eventually add them up.

For one non-vanishing winding number, say $n_1\ne 0$, to solve the BPS condition \eqref{BPS strings} we require $m_1=0$, while $m_2,m_3\in \mathbb{Z}$. Therefore, we have the Schwinger integral 
\begin{equation}
\begin{aligned}
\label{app_1}
a_7^{(1)}&\simeq \frac{2\pi}{\vartheta_{123}} \sum_{n_1\ne 0} \sum_{m_2,m_3\in\mathbb Z} \int_0^\infty \frac{dt}{t^{\frac12}}\, e^{-\pi t\left( n_1^2 \rho_1^2 +\frac{m_2^2}{\rho_2^2} +\frac{m_3^2}{\rho_3^2}\right)}\\
&=\frac{2\pi^2}{3} +4\pi \sum_{n_1>0} \sum_{(m_2,m_3)\ne (0,0)} \frac{e^{ -2\pi n_1 \sqrt{m_3^2 \vartheta_{12}^2 + m_2^2 \vartheta_{13}^2 }}}{ 
    \sqrt{m_3^2 \vartheta_{12}^2 + m_2^2 \vartheta_{13}^2 }}\,,
\end{aligned}
\end{equation}
where the radii are measured in string units and, after Poisson resumming over the Kaluza-Klein momenta, we used \eqref{int exp/t^(3/2)}. 
The contribution of the other two cases, $n_2\ne 0$ and $n_3\ne 0$, is the same up to renaming indices appropriately. 

For two non-vanishing winding numbers, which we can parametrize as $(n_1,n_2)=n(\tilde{n}_1,\tilde{n}_2)$ with $n={\rm gcd}(n_1,n_2)$, the BPS condition \eqref{BPS strings} is a linear Diophantine equation in two variables with solution $m_1 = m \tilde n_2$, $m_2=-m \tilde n_2$, $m\in \mathbb{Z}$. The remaining Kaluza-Klein momentum, $m_3$, is an arbitrary integer $p\in \mathbb{Z}$. 
Therefore, we have the Schwinger integral
\begin{equation}
\label{app_2}
\begin{aligned}
a_7^{(2)}&\simeq \frac{2\pi}{\vartheta_3} \sum_{\tilde n_1, \tilde n_2 \in \mathbb{Z}}\sum_{n \neq 0} \sum_{m,p\in\mathbb{Z}} \int_0^\infty \frac{dt}{t^{\frac12}} e^{-\pi t \left(n^2(\tilde n_1^2 \rho_1^2+\tilde n_2^2 \rho_2^2)+m^2\left(\frac{\tilde n_2^2}{\rho_1^2}+\frac{\tilde n_1^2}{\rho_2^2}\right)+\frac{p^2}{\rho_3^2}\right)}\\
&=-\frac{2\pi^2}{3} +8\pi \sum_{\tilde n_1,\tilde n_2>0}\,  \sum_{n>0} \sum_{(m,p)\ne (0,0)} \frac{e^{ -2\pi n \sqrt{m^2 \vartheta_{12}^2 + p^2 (\tilde n_1^2 \vartheta_{13}^2+\tilde n_2^2 \vartheta_{23}^2) }}}{ \sqrt{m^2 \vartheta_{12}^2 + p^2 (\tilde n_1^2 \vartheta_{13}^2+\tilde n_2^2 \vartheta_{23}^2)}},
\end{aligned}
\end{equation}
where after Poisson resummation over $m,p$ we used \eqref{int exp/t^(3/2)} and \eqref{regcoprime2}. The contribution of the other two cases, $n_1\neq 0 \neq n_3$ and $n_2 \neq 0 \neq n_3$, is the same up to renaming indices appropriately. 

The general case with all three winding numbers being non-vanishing
is a bit more involved, but it can be worked out with the same strategy. The BPS condition \eqref{BPS strings} becomes a homogenoues linear Diophantine equation in three variables, $m_1$, $m_2$, $m_3$, whose solution can be found in appendix \ref{app:Diophantine eqs}. In brief, we can parametrize $(n_1,n_2,n_3,m_1,m_2,m_3) = (n \tilde n_1, n \tilde n_2, n \tilde n_3, \mu_2 X_1 + \mu_1 \tilde n_2/g_2, \mu_2 Y_1-\mu_1 \tilde n_1/g_2, \mu_2 g_2)$, with $\mu_1$, $\mu_2 \in \mathbb{Z}$, $n={\rm gcd}(n_1,n_2,n_3)$ and $g_2 ={\rm gcd}(n_1,n_2)$. In addition, $X_1$ and $Y_1$ are such that $\tilde n_1 X_1 + \tilde n_2 Y_1 + \tilde n_3 g_2=0$. 
Therefore, we have the Schwinger integral
\begin{equation}
\label{app_3}
\begin{aligned}
a_7^{(3)}&\simeq \frac{2\pi}{\vartheta_3} \sum_{n_1,n_2,n_3 \in \mathbb{Z}}\sum_{\substack{m_1,m_2,m_3 \in \mathbb{Z}\\ \\ \sum_{I=1}^3  n_I m_I=0}} \int_0^\infty \frac{dt}{t^\frac12} e^{-\pi t \left(n_1^2 \rho_1^2 + n_2^2\rho_2^2+n_3^2\rho_3^2 + \frac{m_1^2}{\rho_1^2}+\frac{m_2^2}{\rho_2^2}+\frac{m_3^2}{\rho_3^2}\right)}\\
&=\frac{2\pi^2}{3} +16\pi \,\sum_{\tilde n_1, \tilde n_2, \tilde n_3 >0}\,\, \sum_{n>0}\,\, \sum_{\substack{ (m_1,m_2,m_3)\ne (0,0,0)\\ \\ \sum_{I=1}^3 n_I m_I=0}}    \frac{e^{ -2\pi n \sqrt{m_3^2 \vartheta_{12}^2 + m_2^2 \vartheta_{13}^2+m_1^2 \vartheta_{23}^2}}}{\sqrt{m_3^2 \vartheta_{12}^2 + m_2^2 \vartheta_{13}^2+m_1^2 \vartheta_{23}^2} }\,,
\end{aligned}
\end{equation}
where, after Poisson resummation over $\mu_1,\mu_2$, we used \eqref{int exp/t^(3/2)} and \eqref{regcoprime2}.

First we notice that the constant terms arising in all winding
sectors add up to
\begin{equation}
  a_7^{\rm const}=  \frac{2\pi^2}{3}(3-3+1) =\frac{2\pi^2}{3}\,.
\end{equation}  
As for the other terms, it is convenient to organize them according to whether we have single, double and triple world-sheet instantons.
The single instanton receives contributions from both $a_7^{(1)}$ and $a_7^{(2)}$ by setting appropriate summation indices to zero. Say we are interested in the instanton wrapping a 2-cycle of size $\vartheta_{12}$. Then, in \eqref{app_1} we set $m_2=0$ and sum over $m_3\ne 0$ to get 
\begin{equation}
a_7^{(E\!F1)_{12}} \simeq 4\pi \sum_{n_1>0} \sum_{m_3\ne 0} \frac{e^{ -2\pi n_1 |m_3| \vartheta_{12}}}{ |m_3| \vartheta_{12}}\, +\dots.
\end{equation}
From the $n_2\ne 0$ winding sector we get the same contribution, which we take into account by multiplication with a factor of 2. However, there is one more contribution from $a_7^{(2)}$ by setting $p=0$ and summing over $m$ in \eqref{app_2} (and using also $\sum_{\tilde n_1, \tilde n_2>0}1 = -1/2$). The remaining two (with $p\neq 0$) and three winding sectors do not contribute to the single instanton so that in total we get
\begin{equation}
  \label{1instantonstot}
a_7^{(E\!F1)_{12}} \simeq 4\pi \underbrace{(2-1+0)}_{=1} \; \sum_{n>0} \sum_{m\ne 0} \frac{e^{ -2\pi n |m| \vartheta_{12} }}{ |m| \vartheta_{12}}\,,
\end{equation}
where we indicated the contributions from $a_7^{(1)}$, $a_7^{(2)}$ and $a_7^{(3)}$ respectively. 
Similarly, we can collect all contributions to the double instanton case. They arise from $a_7^{(1)}$, $a_7^{(2)}$ with $m_2=0$, i.e.~$\tilde n_2=0$ in \eqref{app_2} and from $a_7^{(3)}$ with $m_1=0$ after using the regularization
\begin{equation}
     \sum_{\substack{\tilde{\mathbf{n}}_3\\
      n_2 m_2+n_3m_3=0\\ m_2,m_3\ne 0 }}  \!\!\!\!\!   1 =-\frac{1}{4}\,,
\end{equation}
which can be derived using \eqref{regcoprime} and \eqref{soldioph2}.
Then, we get
\begin{equation}
  \label{2instantonstot}
a_7^{(E\!F1)_{12}+(E\!F1)_{13}} \simeq 4\pi \underbrace{(1+1 -1)}_{=1}  \sum_{n>0} \sum_{m_2,m_3\ne 0} \frac{e^{ -2\pi n \sqrt{m_3^2 \vartheta_{12}^2 + m_2^2 \vartheta_{13}^2 }}}{ \sqrt{m_3^2 \vartheta_{12}^2 + m_2^2 \vartheta_{13}^2 }}
\end{equation}
where we indicated the contributions from $a_7^{(1)}$, $a_7^{(2)}$ and $a_7^{(3)}$ respectively.  
For the triple instanton case, contributions arise from $a_7^{(2)}$ with a combinatorial factor of $3$, and $a_7^{(3)}$, after using the regularization 
\begin{equation}
\sum_{\substack{\tilde{\mathbf{n}}_3\\ \sum_{I=1}^3  n_I m_I=0\\ m_I\ne 0 }}  \!\!\!\!\!   1 =-\frac12.
\end{equation}
Then, we get
\begin{equation}
  \label{3instantonstot}
\begin{aligned}
a_7^{(E\!F1)_{12}+(E\!F1)_{13}+(E\!F1)_{23}}&=4\pi \underbrace{(0+3 -2)}_{=1} \times \\
&\phantom{aa} \sum_{n>0} \,\sum_{m_1,m_2,m_3\ne 0} \frac{e^{ -2\pi n \sqrt{m_3^2 \vartheta_{12}^2 + m_2^2 \vartheta_{13}^2+m_1^2 \vartheta_{23}^2}}}{\sqrt{m_3^2 \vartheta_{12}^2 + m_2^2 \vartheta_{13}^2+m_1^2 \vartheta_{23}^2} }\,,
\end{aligned}
\end{equation} we indicated the contributions from $a_7^{(1)}$, $a_7^{(2)}$ and $a_7^{(3)}$ respectively.
We notice that these results agree with the ones derived in \cite{Kiritsis:1997em}, eqs.~(5.16), (5.18), (5.20), except for the overall normalization of the triple instanton term.

\section{Linear Diophantine Equations}
\label{app:Diophantine eqs}
Diophantine equations are equations with integer coefficients and for which one is interested in integer solutions. Since BPS conditions appearing in our analysis are of this kind, in this appendix we present a brief overview on linear Diophantine equations and how to solve them. More information can be found for example in the classic reference \cite{Mordell}.

Arguably, the simplest instance is a linear Diophantine equation in two variables, $X$ and $Y$, of the form
\begin{equation}
\label{dioph2var}
a X+ bY = c, \qquad a,b,c \in \mathbb{Z}, \quad \text{$a$ or $b$} \neq 0.
\end{equation}
It is a theorem (e.g.~Theorem 1, chapter 5 of \cite{Mordell}) that this equation admits integer solutions iff gcd$(a,b)$ divides $c$. If $c=0$, the general such solution is $X=Nb$, $Y=-Na$ with $N \in \mathbb{Z}$. If $c\neq 0$, the general solution is 
\begin{equation}
\label{soldioph2}
X = X_0 + Nb\,, \qquad Y = Y_0- Na\,,
\end{equation}
where $(X_0, Y_0)$ is any solution of \eqref{dioph2var}. Indeed, one can check that $aX+ bY - c = a(X-X_0) + b (Y-Y_0) = 0$. 
Notice that one can assume gcd$(a,b)=1$ from the start without loss of generality. Indeed, if this condition is not met, one can divide \eqref{dioph2var} by $g_2\equiv{\rm gcd}(a,b)>1$ to get $\tilde a X + \tilde b Y = \tilde c$, with $\tilde a = a/g_2$ and similarly for $\tilde b$, $\tilde c$. Therefore, one recovers again a linear Diophantine equation with coprime coefficients. The solution in the case $g_2>1$ is given by replacing $a \to a/g_2$ and $b \to b/g_2$ in \eqref{soldioph2}. Since this observation is valid for any number of variables, we will systematically start with coprime coefficients. 
Knowing the solution to the linear Diophantine equation in two variables, one can iterate the algorithm and solve linear Diophatine equations in more variables. 

For our physical applications, namely 1/2-BPS conditions in toroidal compactifications of string/M-theory, it is enough to consider the homogeneous problem. Hence, let us look next at the homogeneous linear Diophantine equation in three variables
\begin{equation}\label{Dioph 3vargen}
aX+bY+cZ=0\,,\quad a,b,c\in\mathbb{Z},\, \quad  \text{$a$ or $b$ or c} \neq 0\,.
\end{equation}
From ${\rm gcd}(a,b,c)={\rm gcd}({\rm gcd}(a,b),c)=1$, we have that ${\rm gcd}(a,b)$ divides not only $cZ=-aX-bY$ but also $Z$. Hence, the solution for $Z$ is
\begin{equation}
Z = M g_2\,,
\end{equation}
with $M \in \mathbb{Z}$. Plugging this into the original expression \eqref{Dioph 3vargen}, the problem is reduced to a non-homogeneous linear Diophantine equation in two variables, whose solution we have already discussed. The most general solution of the equation $a X + b Y=-cg_2M$ is $X=MX_0+N b/g_2$, $Y=MY_0-Na/g_2$, where $(X_0,Y_0)$ is such that $aX_0+bY_0=-cg_2$. Putting everything together, the solution to \eqref{Dioph 3vargen} is then
\begin{align}\label{Dioph 3vargen solution1}
X &= MX_0+N\frac{b}{g_2}\,,\\
Y&=MY_0-N\frac{a}{g_2}\,,\\
\label{Dioph 3vargen solution3}
Z&=Mg_2\,,
\end{align}
with $M,N \in \mathbb{Z}$, and one can check that $a X + b Y +Z =  a(X-MX_0)+ b (Y-MY_0)=0\,$. 
 Once the homogeneous solution is known, the non-homogeneous problem
\begin{equation}
aX+bY+cZ=d
\end{equation}
is solved by
\begin{align}
X &= X_1+MX_0 + N\frac{b}{g_2}\,,\\
Y &= Y_1+MY_0 -N\frac{a}{g_2}\,,\\
Z &= Z_1 + M g_2\,,
\end{align}
where $(X_1,Y_1,Z_1)$ satisfy $aX_1+bY_1+cZ_1=d$. One can check that $aX+bY+cZ-d = a(X-X_1)+b(Y-Y_1)+c(Z-Z_1)=a(MX_0+Nb/g_2)+b(MY_0-Na/g_2)+cMg_2=0$.
Starting from an equation such that $g_3\equiv{\rm gcd}(a,b,c)>1$ amounts to replacing $a\to a/g_3$ and $b \to b/g_3$ in the above solution.

We learned that to solve a homogeneous linear Diophantine equation in $k$ variables we need to know the solution of the non-homogeneous equation with $k-1$ variables. 
By iterating this algorithm, we can solve the homogeneous linear Diophantine equation in four variables
\begin{equation}
aX+bY+cZ+dW=0, \qquad a,b,c,d \in\mathbb{Z} \quad \text{with $a$ or $b$ or $c$ or $d\neq 0$}.
\end{equation}
From gcd$(a,b,c,d)=1$ we have that $g_3$ divides $W$ and thus we can write $W=Lg_3$, with $L\in \mathbb{Z}$.
Denoting $g_3 \equiv {\rm gcd}(a,b,c)$, the full solution is given by
\begin{align}
\label{soldioph4v1}
X &=LX_1+MX_0 + N\frac{b}{g_2}\,,\\
Y &=LY_1+MY_0 -N\frac{a}{g_2}\,,\\
Z &= LZ_1 + M g_2\,,\\
\label{soldioph4v2}
W &= Lg_3\,,
\end{align}
with $aX_1+bY_1+cZ_1=-dg_3$, and one can check that $aX+bY+cZ+dW=a(X-LX_1)+b(Y-LY_1)+c(Z-LZ_1)=a(MX_0+Nb/g_2)+b(MY_0-Na/g_2)+cMg_2=0$. If $g_4={\rm gcd}(a,b,c,d)>1$, we need to replace $a \to a/g_4$, $b\to b/g_4$ in the above solution. Observe that to solve an equation in $k$ variables we need to introduce $k-1$ arbitrary integers.

\subsection*{Momentum matrix in $d=5$}

This algorithm for solving linear Diophantine equations was applied to BPS conditions in different dimensions throughout our analysis. 
To give an explicit example, let us determine the momentum matrix
${\cal M}_{\alpha}^{ij}$  in \eqref{a81loopstw} for $k=5$ ($d=5$) and
the maximal $\alpha=(1,1,1,1,1)$. 
In this case, the BPS condition \eqref{BPS strings} reads
\begin{equation}\label{BPS 5strings}
n_1m_1+n_2m_2+n_3m_3+n_4m_4+n_5m_5=0.
\end{equation}
Assume for simplicity that ${\rm gcd}(n_1,n_2,n_3,n_4,n_5)=1$ and define the following quantities\footnote{In the main text we denoted coprime integers with $\tilde n_i$; here we omit the tilde for convenience.}
\begin{align}
    g_4&={\rm gcd}\left(n_1,n_2,n_3,n_4\right)\,,\\
    g_3&={\rm gcd}\left(\frac{n_1}{g_4},\frac{n_2}{g_4},\frac{n_3}{g_4}\right)\,,\\
    g_2&={\rm gcd}\left(\frac{n_1}{g_3g_4},\frac{n_2}{g_3g_4}\right)\,,\\
    \hat{n}_1&=\frac{n_1}{g_4g_3g_2}\,,\\
    \hat{n}_2&=\frac{n_2}{g_4g_3g_2}\,,\\
    \hat{n}_3&=\frac{n_3}{g_4g_3}\,,\\
    \hat{n}_4&=\frac{n_4}{g_4}\,,
\end{align}
which are necessary to ensure that at each step of solving homogeneous Diophantine equations we are able to have relatively coprime coefficients. The set of particular solutions we will employ is
\begin{align}
    \hat{n}_1X_0+\hat{n}_2Y_0=-\hat{n}_3\,,\\
    g_2(\hat{n}_1X_1+\hat{n}_2Y_1)+\hat{n}_3Z_1=-\hat{n}_4\,,\\
    g_3(g_2(\hat{n}_1X_2+\hat{n}_2Y_2)+\hat{n}_3Z_2)+\hat{n}_4W_2=-n_5\,,
\end{align}
where all factors are integers. Using all of the above to solve \eqref{BPS 5strings}, we find
\begin{align}
m_1 &=\mu_4 X_2 + \mu_3 X_1 + \mu_2 X_0+\mu_1 \hat n_2,\\
m_2 &=\mu_4 Y_2 + \mu_3 Y_1+\mu_2 Y_0 -\mu_1\hat n_1,\\
m_3 &=\mu_4 Z_2 +\mu_3 Z_1+\mu_2 g_2,\\
m_4 &=\mu_4 W_2+\mu_3 g_3 ,\\
m_5 &=\mu_4 g_4 ,
\end{align}
where $\mu_i =\{\mu_1,\mu_2,\mu_3,\mu_4\} \in \mathbb{Z}^4$ are the four arbitrary integers entering the solution of a linear Diophantine equation in five variables. 
Calculating then $\sum_{I=1}^5 m_I^2/\rho_I^2$ and re-expressing it as
$\sum_{i,j=1}^4\mu_i {\cal M}^{ij}_\alpha \mu_j$, we obtain the momentum matrix
\begin{equation}
\label{MMatrix5}
{\cal M}^{ij}_\alpha=\resizebox{!}{.07\textwidth}{$\begin{pmatrix}
\frac{\hat{n}_2^2}{\rho_1^2}+\frac{\hat{n}_1^2}{\rho_2^2}& \frac{\hat{n}_2X_0}{\rho_1^2}-\frac{\hat{n}_1Y_0}{\rho_2^2}&\frac{\hat{n}_2X_1}{\rho_1^2}-\frac{\hat{n}_1Y_1}{\rho_2^2}&\frac{\hat{n}_2X_2}{\rho_1^2}-\frac{\hat{n}_1Y_2}{\rho_2^2}\\
\frac{\hat{n}_2X_0}{\rho_1^2}-\frac{\hat{n}_1Y_0}{\rho_2^2}&
\frac{X_0^2}{\rho_1^2}+\frac{Y_0^2}{\rho_2^2}+\frac{g_2^2}{\rho_3^2}&
\frac{X_0X_1}{\rho_1^2}+\frac{Y_0Y_1}{\rho_2^2}+\frac{g_2Z_1}{\rho_3^2}
&\frac{X_0X_2}{\rho_1^2}+\frac{Y_0Y_2}{\rho_2^2}+\frac{g_2Z_2}{\rho_3^2}\\
\frac{\hat{n}_2X_1}{\rho_1^2}-\frac{\hat{n}_1Y_1}{\rho_2^2}&
\frac{X_0X_1}{\rho_1^2}+\frac{Y_0Y_1}{\rho_2^2}+\frac{g_2Z_1}{\rho_3^2}&\frac{X_1^2}{\rho_1^2}+\frac{Y_1^2}{\rho_2^2}+\frac{Z_1^2}{\rho_3^2}+\frac{g_3^2}{\rho_4^2}&
\frac{X_1X_2}{\rho_1^2}+\frac{Y_1Y_2}{\rho_2^2}+\frac{Z_1Z_2}{\rho_3^2}+\frac{g_3W_2}{\rho_4^2}\\
\frac{\hat{n}_2X_2}{\rho_1^2}-\frac{\hat{n}_1Y_2}{\rho_2^2}
&\frac{X_0X_2}{\rho_1^2}+\frac{Y_0Y_2}{\rho_2^2}+\frac{g_2Z_2}{\rho_3^2}
&\frac{X_1X_2}{\rho_1^2}+\frac{Y_1Y_2}{\rho_2^2}+\frac{Z_1Z_2}{\rho_3^2}+\frac{g_3W_2}{\rho_4^2}
&\frac{X_2^2}{\rho_1^2}+\frac{Y_2^2}{\rho_2^2}+\frac{Z_2^2}{\rho_3^2}+\frac{W_2^2}{\rho_4^2}+\frac{g_4^2}{\rho_5^2}
\end{pmatrix}$}.
\end{equation}
Using the properties of the particular solutions, one can check that ${\rm det}({\cal M}_\alpha)=L_\alpha^2/\vartheta_{(5)}$\,, where $L_\alpha^2=\sum_{I=1}^5n_I^2\rho_I^2$.
Additionally, one can also directly read-off the result for
compactifications on smaller tori. For example, the upper-left
$4\times4$ part is ${\cal M}_\alpha$  for $k=4$, while the upper-left
$3\times 3$ part is ${\cal M}_\alpha$  for $k=3$, and similarly for other dimensions. The structure of the momentum matrix is analogous in compactifications on larger tori.
\newpage 

\bibliography{references}  
\bibliographystyle{utphys}

\end{document}